\documentclass[twocolumn]{aastex631}

\usepackage{upgreek}
\usepackage{subcaption}
\usepackage{CJK}
\usepackage{amsmath}
\newcommand{\yxcc}{y_{\rm X}^{\rm cc}}
\newcommand{\yxia}{y_{\rm X}^{\rm Ia}}
\newcommand{\yfecc}{y_{\rm Fe}^{\rm cc}}
\newcommand{\ymgcc}{y_{\rm Mg}^{\rm cc}}
\newcommand{\yncc}{y_{\rm N}^{\rm cc}}
\newcommand{\yfeIa}{y_{\rm Fe}^{\rm Ia}}
\newcommand{\qxcc}{q_{\rm X}^{\rm cc}}
\newcommand{\qxia}{q_{\rm X}^{\rm Ia}}
\newcommand{\mgfepl}{[\mathrm{Mg/Fe}]_{\rm cc}}

\defcitealias{WAF2017}{WAF}
\defcitealias{weinberg22_apogee}{W22}

\shorttitle{Interpreting multi-element abundances of Quiescent Galaxies}
\shortauthors{Aliza G. Beverage}

\begin{document}

% \title{How can we interpret the chemical composition of quiescent galaxies?\\ Insights from the Abundance Patterns of Milky Way Stars}
% \title{The Abundance Patterns of Massive Quiescent Galaxies across Cosmic Time: \\Insights from a Two-Process Model}

% \title{Interpreting the Chemical Composition of Quiescent Galaxies Over Cosmic Time:\\ Insights from the Abundance Patterns of Milky Way Stars}
\title{Are Nucleosynthetic Yields Universal? Interpreting the Multi-Elemental Abundances of Quiescent Galaxies over Cosmic Time Using Milky Way Stars }

\correspondingauthor{Aliza G. Beverage}
\email{abeverage@carnegiescience.edu}

\author[0000-0002-9861-4515]{Aliza G. Beverage}\thanks{NHFP Hubble Fellow}
\affiliation{Department of Astronomy, University of California, Berkeley, CA 94720, USA}
\affiliation{Observatories of the Carnegie Institution for Science, 813 Santa Barbara Street, Pasadena, CA 91101, USA}
\affiliation{Department of Astrophysical Sciences, Princeton University, 4 Ivy Lane, Princeton, NJ 08544, USA}

\author[0000-0002-6442-6030]{David H. Weinberg}
\affiliation{The Department of Astronomy and Center of Cosmology and AstroParticle Physics, The Ohio State University, Columbus, OH 43210, USA}

\author[0000-0002-7613-9872]{Mariska Kriek}
\affiliation{Leiden Observatory, Leiden University, P.O. Box 9513, 2300 RA Leiden, The Netherlands}

\author[0009-0008-8797-0865]{Nicole Marcelina Gountanis}
\affiliation{Department of Astronomy and Center for Cosmology and AstroParticle Physics, The Ohio State University, Columbus, OH 43210, USA}

\author[0000-0001-7769-8660]{Andrew B. Newman}
\affiliation{Observatories of the Carnegie Institution for Science, 813 Santa Barbara Street, Pasadena, CA 91101, USA}

\author[0000-0002-6442-6030]{Daniel R. Weisz}
\affiliation{Department of Astronomy, University of California, Berkeley, CA 94720, USA}

\begin{abstract}

The detailed abundance patterns of quiescent galaxies offer powerful constraints on their formation and evolution. Yet physical insight remains elusive, as nucleosynthetic yields are notoriously uncertain. We introduce a framework that circumvents this problem by using Milky Way abundance trends as empirical proxies for the yields. Applied to quiescent galaxies spanning three redshifts, SDSS ($z\sim0$), LEGA-C ($z\sim0.7$), and \textit{JWST}/SUSPENSE ($z\sim2$), our approach recovers the $\alpha$- and Fe-peak abundances with a median offset of $\sim$0.05~dex across 14 elements, compared to $\sim$0.23~dex for theoretical yields. The largest discrepancies arise in N, Sr, Ba, and (at $z\sim2$) C, all of which depend on AGB enrichment, a channel we do not explicitly model. We explore the impact of a top-heavy IMF on our predictions and find that it can shift the IMF-averaged core-collapse supernova yields by $\sim$0.05--0.2~dex in a direction that reduces the overall residuals. Surprisingly, the predictions succeed even without modeling the full chemical-evolution history of a galaxy; just Mg and Fe, which trace the relative contributions of core-collapse and Type Ia supernovae, suffice to predict $\alpha$- and Fe-peak elements. The success of the empirical yields, previously demonstrated in dwarf galaxies and the Milky Way disk, and now extended to massive quiescent galaxies, suggests that $\alpha$- and Fe-peak nucleosynthetic yields are largely universal. This lack of complexity makes galaxy abundance patterns highly predictable. Embedding these empirical yields in SPS models will improve inferences on stellar population properties and star formation histories. Moreover, incorporating them into cosmological simulations will produce more observationally motivated predictions.

\end{abstract}


% \linespread{1.8}

\section{Introduction} \label{sec:intro}

% rybizki+17:
% In the past, theoretical yields have produced mismatches with the observations,
% leading to the concept of “empirical yields” (François et al.
% 2004; Henry et al. 2010). Yet, many abundance trends are not
% reproduced (Argast et al. 2002; Kobayashi et al. 2006) and the
% physical shortcomings of stellar nucleosynthetic yield models
% are still under debate (Nomoto et al. 2013; Fink et al. 2014;
% Pignatari et al. 2016; Müller 2016).

Chemical abundances encode detailed records of a galaxy's star formation, quenching, and assembly history \citep[e.g.,][]{tinsley_stellar_1979, matteucci_abundance_1994, thomas_epochs_2005, choi_assembly_2014, kriek_massive_2016, beverage_elemental_2021}. Since each element arises from distinct nucleosynthetic processes associated with stars of different masses and lifetimes, multi-elemental abundances additionally carry information about the origins of chemical elements and the initial mass function \citep[IMF;][]{conroy_early-type_2014, denbrok_recovery_2024}.

Extracting this information is challenging. On the observational side, chemical information is buried in faint stellar absorption features, requiring high-S/N continuum spectroscopy and sophisticated full-spectrum fitting techniques. Improvements in stellar population models and full-spectrum fitting \citep[e.g.,][]{conroy_counting_2012, villaume_extended_2017, conroy_metal-rich_2018} have yielded precise multi-element abundance patterns for thousands of nearby galaxies \citep[e.g.,][]{zhuang_glimpse_2023, conroy_early-type_2014}, and advancements in ground-based instrumentation, along with the advent of \textit{JWST} have pushed these measurements to unprecedented distances, from $z\sim0.7$ with LEGA-C \citep{van_der_wel_vlt_2016, van_der_wel_large_2021, beverage_carbon_2023} and out to $z\sim3$ using \textit{JWST} \citep{jafariyazani_chemical_2024, beverage_suspense_2025}.

The deeper challenge remains the interpretation of these chemical abundance measurements. Translating multi-element abundances into physical insight requires nucleosynthetic yields, which are notoriously uncertain. They are difficult to measure directly from supernova explosions and hard to predict theoretically, due to uncertainties in stellar mass loss, rotation, supernova explosion mechanisms, and black hole formation \citep[e.g.,][]{nomoto_nucleosynthesis_2013}. A striking example is the [O/Mg] problem: most theoretical yield tables under-predict observed Mg abundances by up to a factor of four \citep[e.g.,][]{thoams1998_paper1,griffith2021}. 

These observational and modeling challenges have historically limited chemical studies of massive quiescent galaxies to just two elements: Mg and Fe. Magnesium traces rapid enrichment from core-collapse supernovae (CCSNe), while Fe additionally traces delayed enrichment from Type Ia supernovae (SNe~Ia), so [Mg/Fe] serves as a rough proxy for star-formation timescale without requiring explicit chemical evolution modeling or the use of nucleosynthetic yields. Despite its simplicity, the relative [Mg/Fe] ratio has produced critical insights: massive low-redshift galaxies formed rapidly and early, indicated by their elevated [Mg/Fe] ratios and old stellar populations \citep[e.g.,][]{trager_stellar_2000, thomas_epochs_2005, mcdermid_atlas3d_2015}. At higher redshifts, massive quiescent galaxies have even higher [Mg/Fe] ratios, posing fundamental questions about the evolutionary paths of massive galaxies across cosmic time \citep[][]{zhuang_glimpse_2023, carnall_jwst_2024, beverage_heavy_2024, beverage_suspense_2025}.

While multi-element abundance measurements beyond Mg and Fe are now possible across cosmic time \citep[][]{conroy_early-type_2014, jafariyazani_resolved_2020, beverage_carbon_2023, beverage_suspense_2025}, interpreting them requires better yields, and few studies have attempted to do so. Notable exceptions include \citet{conroy_strontium_2013}, who used s-process elements (Sr, Ba) from AGB stars to sharpen chronometric constraints on star-formation histories, and \citet{conroy_early-type_2014}, who showed that Ca tracks Fe rather than typical $\alpha$-elements. More recently, \citet{denbrok_recovery_2024} attempted to constrain the high-mass IMF using theoretical yields but found that chemical evolution models fail to reproduce multi-element abundance patterns, a direct consequence of the uncertainty in yields.

One promising path forward is to use the Milky Way as a calibration source for nucleosynthetic yields. \citet{weinberg_chemical_2019} and \citet{griffith2021} show that the multi-element abundance trends in the APOGEE survey \citep[][]{majewski2017} are remarkably uniform throughout the Milky Way disk, and can be decomposed into two components: one tracing prompt enrichment from CCSNe, and one tracing delayed enrichment from SNe~Ia. This ``two-process model'' effectively provides an empirical, IMF-averaged calibration of nucleosynthetic yields directly from observations, bypassing the need for uncertain theoretical predictions.

In this paper, we ask whether the integrated abundance patterns of massive quiescent galaxies follow the same trends with $[\mathrm{Mg/H}]$ and $[\mathrm{Mg/Fe}]$ as Milky Way stars. We test this by implementing empirically calibrated yields within chemical evolution models tuned to match the observed $[\mathrm{Mg/H}]$ and $[\mathrm{Mg/Fe}]$ of massive quiescent galaxies across cosmic time. Section~\ref{sec:data} describes the observational data. Section~\ref{sec:model} describes the empirical yields and modeling approach. We present results in Section~\ref{sec:results} and discuss implications in Section~\ref{sec:discussion}. Section~\ref{sec:conclusion} summarizes our findings.

% The connection between the two-process model and IMF-averaged yields is nuanced in the case of metallicity-dependent yields or elements with a large contribution from asymptotic giant branch (AGB) stars (see Section~\ref{sec:subtle} below).

% In this paper, we test whether the abundance patterns of massive quiescent galaxies can be reproduced by implementing these empirically calibrated yields within galactic chemical evolution models tuned to match their observed [Mg/H] and [Mg/Fe]. The connection between the 2-process model and IMF-averaged yields is nuanced in the case of metallicity-dependent yields or elements with a large contribution from asymptotic giant branch (AGB) stars (see Section~\ref{sec:subtle} below). At a more directly empirical level, we assess whether the integrated abundance patterns of massive quiescent galaxies follow the same trends with [Mg/H] and [Mg/Fe] observed in Milky Way stars. Section 2 describes the observational data used in this analysis. In Section 3, we describe the empirical yields and our method of applying them to massive quiescent galaxies. We present the results in Section 4, and in Section 5, we discuss their implications for galaxy evolution. Finally, Section 6 summarizes our findings and concludes the paper.

\section{Data}\label{sec:data}

In this work, we utilize multi-element abundance measurements of massive quiescent galaxies from three spectroscopic surveys spanning different cosmic epochs: SDSS \citep[$z\sim0$;][]{conroy_early-type_2014, beverage_carbon_2023}, LEGA-C \citep[$z\sim0.7$;][]{beverage_carbon_2023}, and \textit{JWST}-SUSPENSE \citep[$z\sim2$;][]{slob_jwst-suspense_2024, beverage_suspense_2025}. These samples probe comparable velocity-dispersion ranges, although the SDSS data extend to significantly lower values. Below, we summarize the galaxy selection criteria and abundance measurement techniques for each sample.

The selection and abundance measurements for the SDSS sample ($0.025 <z< 0.06$) were first presented in \citet{conroy_early-type_2014}. Galaxies were chosen to be quiescent, requiring no detectable H$\alpha$ or [O\textsc{II}]$\lambda$3727 emission, and to have early-type morphologies, defined by their location on the fundamental plane \citep{graves_dissecting_2010}. The corresponding SDSS spectra were binned by velocity dispersion into seven stacks, each containing about 1000 galaxies on average. Spectra were continuum-normalized and smoothed to a velocity dispersion of $350\,{\rm km\,s^{-1}}$ before fitting. In this study, we use the updated abundance measurements from \citet{beverage_carbon_2023} for these stacked spectra.

\begin{figure*}
    \centering
    \includegraphics[width=1\textwidth]{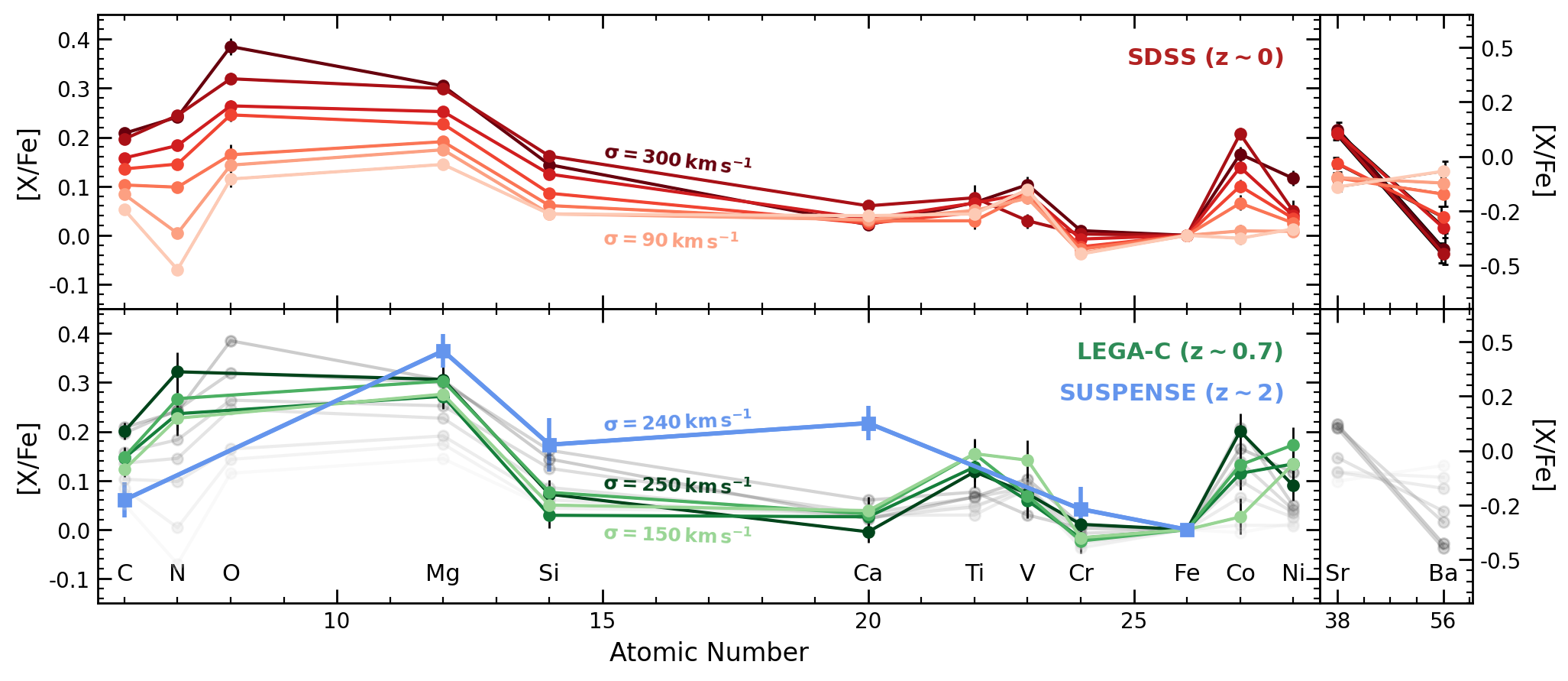}
    \caption{Multi-element abundance patterns used in this study. In the top row, we show the abundance ratios $[X/\mathrm{Fe}]$ for the low-redshift (SDSS, $z\sim0$) galaxy sample selected by \citet{conroy_early-type_2014}, with abundance measurements from \citet{beverage_carbon_2023}. Darker line colors indicate galaxies in higher velocity dispersion ($\sigma$) bins. The small right-hand panel displays measurements for heavier elements (Sr, Ba). In the bottom panel, we present abundance ratios for intermediate-redshift (LEGA-C, $z\sim0.7$; green lines) and high-redshift (SUSPENSE, $z\sim2$; blue lines), with gray lines showing the SDSS sample for comparison. Error bars indicate typical measurement uncertainties for each element.}

    \label{fig:data}
\end{figure*}

The $z\sim0.7$ sample is presented in \citet{beverage_carbon_2023}. It is drawn from the third data release of LEGA-C \citep{van_der_wel_vlt_2016, straatman_large_2018, van_der_wel_large_2021}. LEGA-C is a deep spectroscopic survey of nearly 4000 massive galaxies at $0.6\lesssim z\lesssim 1.0$ in the COSMOS footprint conducted with VIMOS on the VLT. The \citet{beverage_carbon_2023} quiescent galaxies were selected using the $UVJ$ quiescent criteria of \citet{muzzin_public_2013}. Galaxies were also selected to have a signal-to-noise ratio S/N$>15$ \AA$^{-1}$ in their rest-frame optical spectra and to cover key Balmer, Mg and Fe absorption features. This sample of 135 galaxies was divided into four velocity-dispersion bins, with on average 40 galaxies per bin. Each spectrum was fit individually to avoid complications from continuum normalization and stacking. After fitting, posterior distribution functions within each bin were combined to emulate stacked results (see \citealt{beverage_carbon_2023} for more details).

The $z\sim2$ sample is from the Cycle 1 \textit{JWST}-SUSPENSE survey (ID:2110), which obtained ultradeep (16.4 hr) NIRSpec-MSA/G140M-F100LP observations of 20 massive quiescent galaxies at $z=1-3$ in the COSMOS footprint \citep{slob_jwst-suspense_2024}.  Quiescent galaxies were selected using the same $UVJ$ criteria as for LEGA-C. As with LEGA-C, the posterior distributions of the 16 galaxies with individual abundance constraints were combined after fitting, yielding a stacked abundance pattern that reflects the median properties of the sample. For more details, we direct readers to \citet{beverage_suspense_2025}. The {\it JWST} data used in this paper can be found in MAST: \dataset[10.17909/6wjp-qb35]{https://doi.org/10.17909/6wjp-qb35}.

The elemental abundance patterns of the galaxies in all three samples were modeled using a full-spectrum approach, using the \texttt{absorption line fitter (alf)}. \citep{conroy_counting_2012, conroy_metal-rich_2018, alfsoft2023}. These models rely on metallicity-dependent MIST isochrones \citep{choi_mesa_2016} and the MILES and IRTF empirical stellar libraries \citep{sanchez-blazquez_medium-resolution_2006, villaume_extended_2017}, along with synthetic metallicity- and age-dependent elemental response spectra for 19 elements \citep[][]{conroy_metal-rich_2018}. All three redshift samples were fit using the same settings; however, the $z\sim2$ sample was fit using \texttt{alf$\alpha$} \citep{beverage_suspense_2025}, a Python implementation of \texttt{alf}. The primary difference between these fitting codes is the posterior sampling; the original \texttt{alf} code uses MCMC, whereas the updated Python version (\texttt{alf}$\alpha$) uses a dynamic nested sampler \citep[\texttt{dynesty};][]{speagle_dynesty_2020}. We note that these two fitting approaches provide similar results.

While the code fits 19 elements, constraints are available only for a subset of these elements in each sample, depending on the wavelength coverage and spectral S/N. For the $z\sim0.7$ and $z\sim2$ samples, elements were considered constrained if their posterior distributions were distinguishable from the input (uniform) prior \citep[see][]{beverage_carbon_2023, beverage_suspense_2025}. Additionally, for the $z\sim0.7$ sample, model-recovery tests were performed to further validate the constraints. In the end, SDSS has reliable abundances for 14 elements, LEGA-C for 11, and the $z\sim2$ SUSPENSE sample for 6. 

In Figure~\ref{fig:data}, we summarize the multi-element abundance patterns for each sample, illustrating which elements are well-constrained at each redshift. The top panel shows abundance measurements for SDSS galaxies, stacked by velocity dispersion. The bottom panel shows the LEGA-C abundance patterns in green for 4 velocity dispersions bins (150, 180, 210, 250~km$\,$s$^{-1}$) and SUSPENSE abundance pattern in blue for a single velocity dispersions bin (240~km$\,$s$^{-1}$), with the SDSS measurements shown in gray for comparison.

\section{Predicting the multi-element abundances of quiescent galaxies}
\label{sec:model}
The goal of this study is to test whether empirical nucleosynthetic yields derived from the Milky Way apply to massive quiescent galaxies. We first describe how these yields are derived, then how they are applied to the observations.

\subsection{Nucleosynthetic yields from Milky Way stars}
\label{sec:yields}
The APOGEE survey \citep[][]{majewski2017} measured abundances of 15-20 elements in hundreds of thousands of stars in the Milky Way disk, bulge, and halo. \citet[][hereafter W22]{weinberg22_apogee} show that a star's APOGEE abundances can be predicted with surprising accuracy from its [Mg/H] and [Mg/Fe] alone, with rms scatter of $\sim$0.01-0.05 dex for most elements. The physical origin of this low dimensionality is that chemical enrichment is dominated by two nucleosynthetic channels with distinct timescales: a prompt process associated with massive stars and CCSNe, and a delayed process dominated by SNe Ia. As a result, a star’s chemical abundance pattern can be understood as a linear combination of the abundance patterns produced by these two channels, with the abundance of element X expressed as 
\begin{equation}
\label{eq:xh}
[\mathrm{X/H}]= \log_{10}[ A_\mathrm{cc}\;q_{\mathrm{cc}}^X + A_{\mathrm Ia}\;q_{\mathrm{Ia}}^X],
\end{equation}
where $A_{\rm cc}$ and $A_{\rm Ia}$ represent the relative contributions of the prompt (CCSN) and delayed (SNe~Ia) enrichment channels, respectively, and $q_{\rm cc}^X$ and $q_{\rm Ia}^X$ describe the abundance patterns associated with each channel. \citetalias{weinberg22_apogee} refer to $q_{\rm cc}^X$ and $q_{\rm Ia}^X$ as the two-process vectors; as we describe in Section~\ref{sec:subtle}, these can be converted into quantities that effectively serve as empirically calibrated, IMF-averaged nucleosynthetic yields.

Using the large sample size and high precision of APOGEE, \citetalias{weinberg22_apogee} empirically derived these effective yields, $q_{\rm cc}$ and $q_{\rm Ia}$. In this work, we re-derive their values following the same methodology. We briefly summarize the procedure and refer the reader to Section~2 of \citetalias{weinberg22_apogee} for a full derivation. 

The method relies on the following assumptions:
\begin{enumerate}
    \item Gas enrichment is dominated by two nucleosynthetic channels: CCSNe and SNe~Ia.
    \item Mg is produced exclusively by CCSNe.
    \item The CCSN and SNIa yields of Mg and Fe, $q_{\rm cc}$ and $q_{\rm Ia}$, are independent of metallicity.
\end{enumerate}

To begin, the Milky Way disk stars are split into high-Ia (low-$\alpha$) and low-Ia (high-$\alpha$) sequences in the $[\mathrm{Fe/H}]$--$[\mathrm{Mg/Fe}]$ plane. The left panel of Figure~2 shows the APOGEE data with this division, along with the two median sequences derived by \citetalias{weinberg22_apogee}. This division distinguishes stars with substantial SNe~Ia enrichment from those dominated by CCSNe.
% allowing the contributions from the prompt and delayed enrichment channels to be isolated.

Next, we quantify the relative contributions of CCSNe and SNe~Ia at each point along these median sequences. Because Mg is produced purely by CCSNe, it provides a direct proxy for the CCSN contribution, $A_{\rm cc}$:
\begin{equation}
\label{eq:acc}
A_{\rm cc} = 10^{\rm [Mg/H]} .
\end{equation}

The contribution from SNe~Ia is inferred instead from the depletion of $[\mathrm{Mg/Fe}]$ relative to its CCSN-only value. In the absence of SNe~Ia enrichment, $[\mathrm{Mg/Fe}]$ equals the CCSN plateau, $[\mathrm{Mg/Fe}]_{\rm cc}$. As SNe~Ia contribute Fe (but no Mg), [Mg/Fe] decreases below this plateau. The contribution from SNe Ia is therefore given by
\begin{equation}
\label{eq:aia}
A_{\rm Ia} = A_{\rm cc}
\frac{10^{\rm -[Mg/Fe]} - 10^{-[\rm Mg/Fe]_{\rm cc}}}
{1 - 10^{-[\rm Mg/Fe]_{\rm cc}}} .
\end{equation}
Hence, $A_{\rm Ia} = A_{\rm cc}=1$ for a star with solar abundances.

After determined $A_{\rm cc}$ and $A_{\rm Ia}$ along the median low-Ia and high-Ia sequences in the Milky Way from Eqs.\;\ref{eq:acc} and \ref{eq:aia}, we can then solve for $q_{\rm cc}$ and $q_{\rm Ia}$ as a function of metallicity using Eq.~\ref{eq:xh}. At each metallicity, the low-Ia and high-Ia sequences provide two independent combinations of $(A_{\rm cc}, A_{\rm Ia})$ and their corresponding abundances [X/H]. These combinations define a system of two linear equations, which can be solved independently at each metallicity and for each element $X$ to obtain $q_{\rm cc}^X(\rm{[Mg/H]})$ and $q_{\rm Ia}^X(\rm{[Mg/H]})$.

A key ingredient in Eq.~\ref{eq:aia} is the choice of the CCSN plateau, $[\mathrm{Mg/Fe}]_{\rm cc}$. \citetalias{weinberg22_apogee} adopts $[\mathrm{Mg/Fe}]_{\rm cc} = 0.3$, while in this work we adopt a much higher value of $[\mathrm{Mg/Fe}]_{\rm cc} = 0.65$.  \citet{gountanis2025} demonstrated that lower values of [Mg/Fe]$_{\rm cc}$ are unable to reproduce the high $[\mathrm{Mg/H}]$ observed in nearby quiescent galaxies. Such a high CCSN plateau is also motivated by the observed [Mg/Fe] in the Milky Way's lowest-metallicity stars \citep[e.g.,][]{Cohen2004,conroy22_h3}. 

The choice of $\mgfepl$\ has a large impact on $\qxcc$\ and $\qxia$\ individually, but it has no effect on the predicted galaxy abundances. This is because $\mgfepl$\ enters the calculation twice, in the same way: first when deriving the yields from Milky Way stars, and again when predicting galaxy abundances (see Section~\ref{sec:mod_app_to_obs}). In both cases, it sets the reference level used to infer the SNe~Ia contribution from an observed [Mg/Fe]. Because the same reference is used in both steps, any shift in $\mgfepl$\ changes the yields and the galaxy prediction by the same amount, and the effect cancels exactly \citep[][]{griffith24, sit25}.

For most elements, we derive $q_{\rm cc}$ and $q_{\rm Ia}$ using the high-Ia and low-Ia sequences from the APOGEE survey analysis \citep{weinberg22_apogee}. However, because Ti, Sr, and Ba are not included in the original APOGEE element set, we instead use Ba and Ti measurements from \citet{griffith19_galah} based on GALAH data, and the GALAH Y measurements as a proxy for Sr. For C and N, we use the high- and low-Ia sequences from \citet{roberts_nature_2024}, who used APOGEE subgiant stars (unaffected by dredge-up), providing a snapshot of their \textit{natal} chemical compositions. Our adopted $q_{\rm cc}$ and $q_{\rm Ia}$ are shown as a function of metallicity in the appendix Figure~\ref{fig:qs}.

\subsection{Empirical IMF-averaged yields}
\label{sec:subtle}

The two-process vectors, $\qxcc$\ and $\qxia$, represent the effective IMF-averaged yields, but converting them to actual IMF-averaged yields, $\yxcc$\ and $\yxia$, enables direct use in chemical evolution calculations and cosmological simulations. $y_{\rm X}$ is defined to be the mass of element X produced per unit mass of stars formed. For elements whose yields are metallicity-independent, we can recover the yield of element X from $\qxcc$\ and $\qxia$\ by scaling them by the Mg and Fe yield and the appropriate solar abundance ratio:

\begin{equation}
\label{eq:yxcc}
    \yxcc = \ymgcc \cdot \frac{\qxcc}{q_{\rm Mg}^{\rm cc}}\left(\frac{Z_{\rm X}}{Z_{\rm Mg}}\right)_\odot
\end{equation}

\begin{equation}
\label{eq:yxia}
    \yxia = \yfeIa \cdot \frac{\qxia}{q_{\rm Fe}^{\rm Ia}}\left(\frac{Z_{\rm X}}{Z_{\rm Fe}}\right)_\odot.
\end{equation}

Here $\rm \left(X/Mg\right)_\odot=10^{[X/Mg]}$ and $\rm \left(X/Fe\right)_\odot=10^{[X/Fe]}$ are solar abundance ratios on a linear scale. These equations assume that the Mg and Fe yields are also independent of metallicity. $\ymgcc$ and $\yfeIa$ are the IMF-averaged Mg and Fe yields, which set the overall mass scale of the yields. We specify these yields following the approach of \citet{weinberg24_yields} and \citet{gountanis2025}. For CCSNe, we adopt the mean Fe yield measured observationally by \citet{Rodriguez23_SNe}, which implies 
\begin{equation}
\label{eq:yccfe}
\frac{y^{\rm{cc}}_{\rm{Fe}}}{Z_{\rm{Fe,\odot}}}=0.345,
\end{equation}
where $y^{\rm{cc}}_{\rm{Fe}}$ is the mass of iron produced per unit mass of
star formation. The mean Mg yield then follows directly from our CCSN plateau, $[\mathrm{Mg/Fe}]_{\rm cc}=0.65$ that we defined in Section~\ref{sec:yields},
\begin{equation}
\label{eq:yccmg}
\frac{y^{\rm{cc}}_{\rm{Mg}}}{Z_{\rm{Mg,\odot}}}=1.54. %1.22.
\end{equation}

For SNe~Ia, we calibrate the Fe yield by asking: what Fe yield is required to explain the observed decline in [Mg/Fe] from the CCSN plateau to its value in the solar neighborhood today ([Mg/Fe]$=0$), assuming a smooth, extended Milky Way SFH? Following the derivation in \citet{weinberg24_yields}, this requirement implies an IMF-averaged Fe yield of
\begin{equation}
\label{eq:yiafe}
\frac{y^{\rm{Ia}}_{\rm{Fe}}}{Z_{\rm{Fe,\odot}}}=0.967,
\end{equation}
given our values of $\ymgcc$ and $\yfecc$.

Combining Eqs.~\ref{eq:yccmg}-\ref{eq:yiafe} with our adopted solar abundances gives
\begin{equation}
\begin{aligned}
\{\ymgcc,\yfecc,\yfeIa\} = 
&\{10.795,\\
&\;\;4.730, 14.907\} \times 10^{-4}.
\end{aligned}
\label{eqn:yvals}
\end{equation}

\noindent Altogether, Eqs.~\ref{eq:yxcc} and \ref{eq:yxia}, combined with Eq.~\ref{eqn:yvals}, provide the IMF-averaged yields of CCSN and SNIa. The larger values of $\ymgcc$ and $\yfeIa$ relative to those of \citet{weinberg24_yields} reflect our adoption of a higher [Mg/Fe]$_\mathrm{cc}$. We provide these empirically calibrated yields in Tables~\ref{tab:yields} and \ref{tab:yields_cnba}.

The above derivation assumes metallicity-independent yields, which are approximately true for most $\alpha$ and Fe-peak elements. However, if $\qxcc$\ or $\qxia$\ depend on metallicity, then the conversion to yields is less straightforward---the abundance ratios of Milky Way stars at some metallicity will reflect the yields of stars that formed earlier, at a different, typically lower, metallicity. We ignore this complication in the paper and simply use $\qxcc$ and $\qxia$ in Equations \ref{eq:yxcc} and \ref{eq:yxia} to infer $\yxcc$ and $\yxia$. More accurate empirical yields for elements with metallicity-dependent $\yxcc$ or $\yxia$ will require more sophisticated models of observed abundance patterns \citep[e.g.,][]{sanders2025}. That said, as we will show, $\yxcc$\ and $\yxia$\ successfully predict $\alpha$ and Fe-peak abundance ratios across a wide range of galactic environments, and thus this metallicity-independent approximation appears to be a reasonable one. 

The relation of the two-process model to yields is also more complicated for an element whose delayed contribution arises from AGB stars rather than from SNIa. \citetalias{weinberg22_apogee} found that the two-process model predicts the abundances of C, N, and Ce with reasonable accuracy, presumably because stars with a large contribution of Fe from delayed SNIa also have a larger contribution from delayed AGB enrichment. \citet{griffith_residual_2022} found similar results for Y and Ba, though they found correlated deviations of Y and Ba that could be explained by a three-process model with a distinct AGB component. Because the characteristic time delays of SNIa and AGB enrichment are different, their relative contributions to a given stellar population will depend on the star formation history. Given the sharp differences in SFH between the Milky Way disk and massive quiescent galaxies, we might expect our Milky Way calibrated yields to give less accurate predictions for AGB elements in quiescent galaxies. In our study, the well-measured elements that we expect to have large AGB contributions are C, N, Sr, and Ba.

\begin{figure*}
    \centering
    \includegraphics[width=1\textwidth]{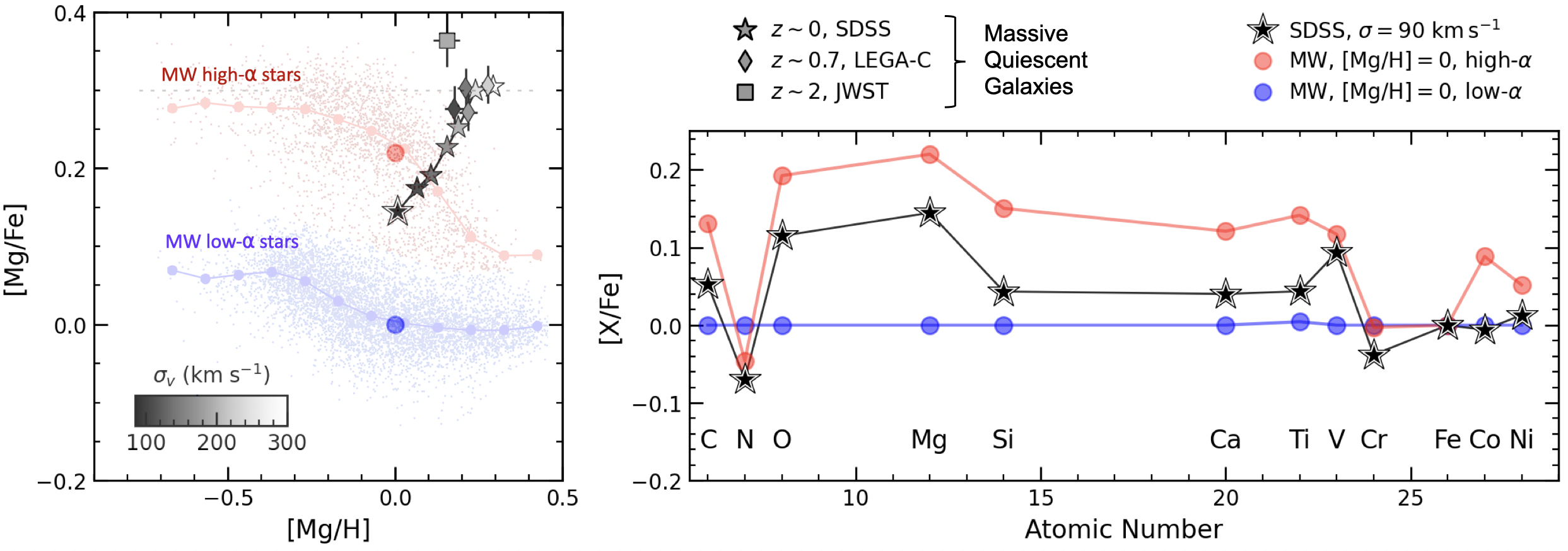}
    \caption{\textit{Left:} The [Mg/Fe] vs. [Mg/H] of Milky Way stars (small circles) from APOGEE \citep{weinberg22_apogee} compared with typical values for massive quiescent galaxies at $z\sim0$ (SDSS; stars), $z\sim0.7$ (LEGA-C; diamonds), and $z\sim2$ (\textit{JWST}-SUSPENSE; square). The quiescent galaxies are color-coded by their velocity dispersions. Large circles and the highlighted star mark populations at ${\rm [Mg/H]=0}$.The APOGEE stars are separated into high-$\alpha$ (red) and low-$\alpha$ (blue) sequences. \textit{Right:} The abundance patterns of a typical solar-metallicity star on the high-$\alpha$ (red) and low-$\alpha$ (blue) sequences. The difference between these abundance patterns can be explained by the low-$\alpha$ star being formed from gas with more SNe Ia enrichment. Black stars show the abundance pattern of a typical solar-metallicity quiescent galaxy from SDSS. Such galaxies are intermediate in [Mg/Fe] between high-$\alpha$ and low-$\alpha$ Milky Way stars with [Mg/H]=0 (left panel), and their abundance pattern is mostly intermediate between that of such stars in other elements (right panel).
    % This idea, that the abundance patterns simply reflect a linear combination of the SNe Ia and CCSN yield patterns, is used to derive the nucleosynthetic yields and to predict chemical abundance patterns of galaxies (Section~\ref{sec:model}. In this panel, we also show how the abundance pattern of a solar-metallicity SDSS galaxy compares to those in the MW.
    }
    \label{fig:apogee_vs_galaxies}
\end{figure*}

\subsection{Application to observations}
\label{sec:mod_app_to_obs}
With the empirically-derived yields in hand, we can predict the complete chemical abundance patterns of massive quiescent galaxies across cosmic time. We present two complementary approaches. In the simplest case, we apply yields at a single metallicity, ignoring any effect of metallicity-dependent yields. In the second case, we incorporate a chemical evolution model to estimate each galaxy’s metallicity history.\footnote{We will show that these two methods usually lead to similar results.}

\begin{figure*}
    \centering
    \includegraphics[width=1\linewidth]{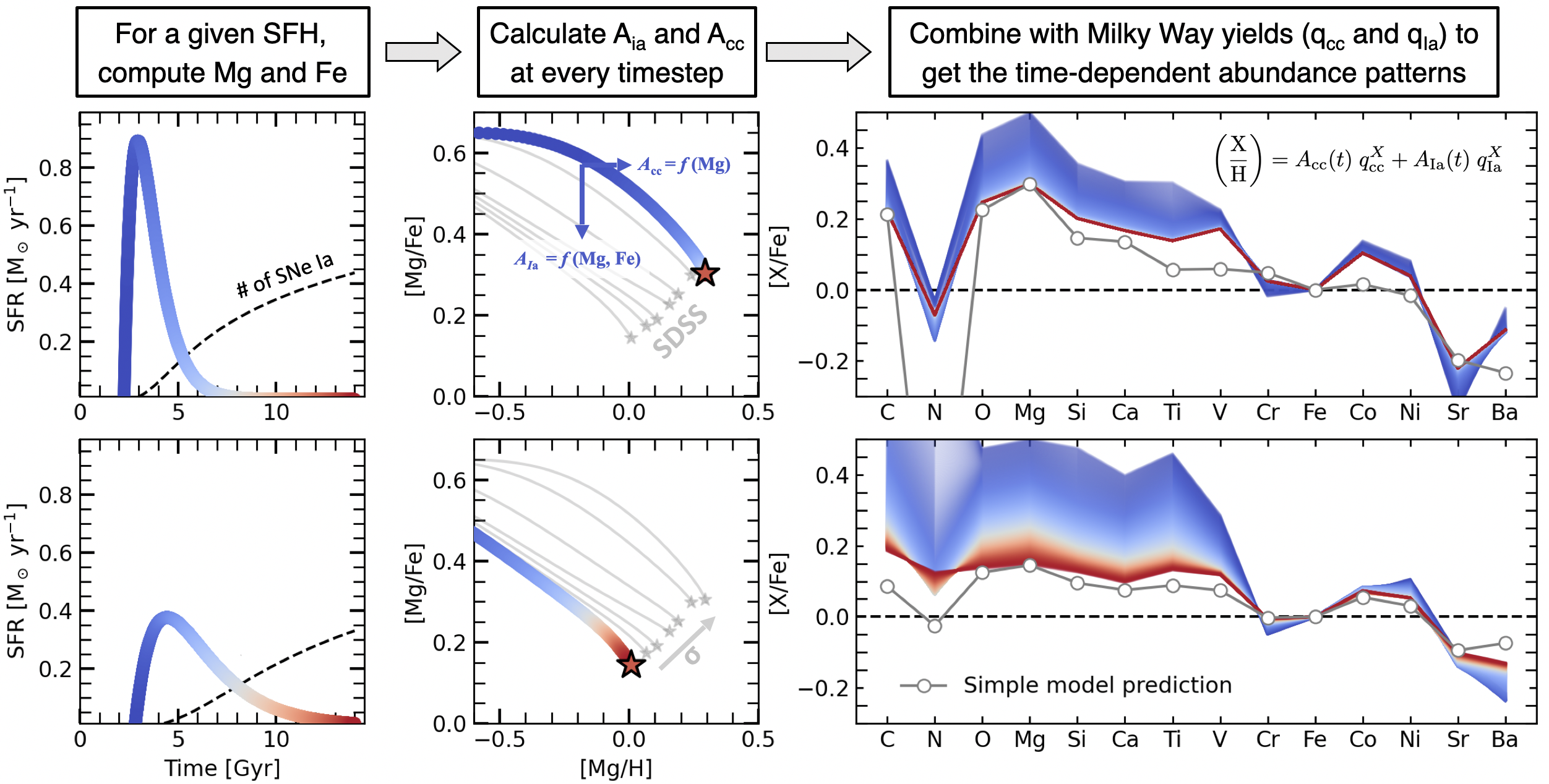}
    \caption{Illustration of our chemical evolution modeling framework applied to the highest (top) and lowest (bottom) velocity dispersion bins of the SDSS sample. Left: Best-fit SFHs and (in black) the cumulative number of SNe~Ia. The high-dispersion galaxy has a shorter SFH and therefore experiences fewer SNe~Ia, consistent with its higher $[\mathrm{Mg/Fe}]$. Middle: The best-fit light-weighted $[\mathrm{Mg/Fe}]$ vs.\ $[\mathrm{Mg/H}]$ tracks color-coded by time, for all seven SDSS bins (lines). Observed values are shown as stars. Right: Light-weighted multi-element abundance patterns as a function of time, color-coded from early (blue) to late (red). The final time step (red) represents the predicted observable stellar abundances. We also show the predicted abundance pattern from the basic application of our empirically-derived yields (Section~\ref{sec:mod_app_to_obs}; gray lines), which anchors the abundances to the observed $[\mathrm{Mg/H}]$ and $[\mathrm{Mg/Fe}]$ without modeling the full enrichment history.}
    \label{fig:model}
\end{figure*}

Figure~\ref{fig:apogee_vs_galaxies} illustrates the first approach. The left panel shows the distribution of APOGEE stars in the [Mg/Fe]--[Mg/H] plane, along with the high-$\alpha$ (red) and low-$\alpha$ (blue) median sequences. Overplotted as gray points are the measured [Mg/H] and [Mg/Fe] of our galaxy samples (stars: $z\sim0$; diamonds: $z\sim0.7$; square: $z\sim2$).

\begin{figure*}
    \centering
    \includegraphics[width=1\textwidth]{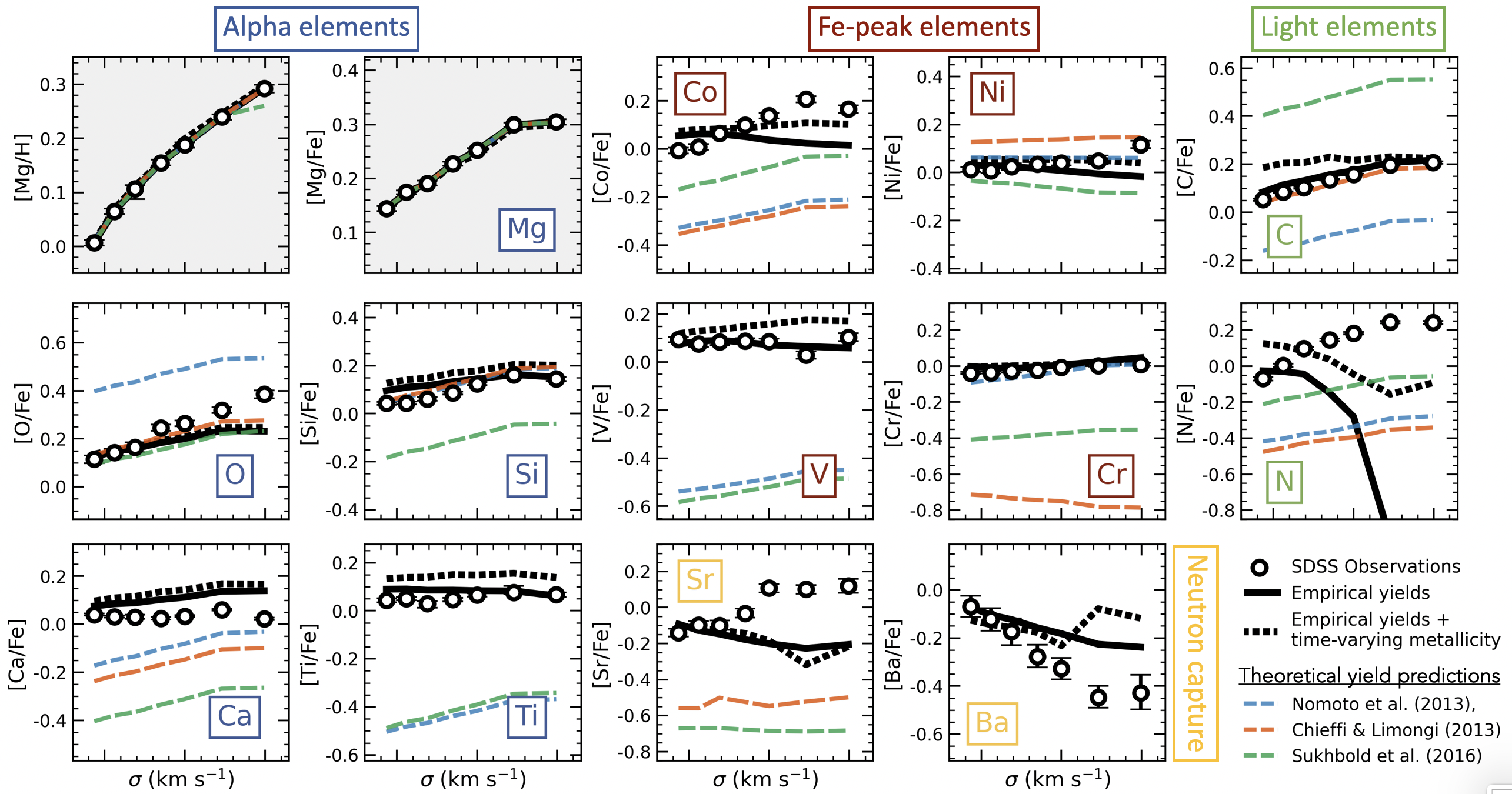}
    \caption{Comparison between observed (circles) and predicted (lines) abundance ratios [X/Fe] of the SDSS sample as a function of velocity dispersion. All predictions are calibrated to the observed [Mg/H] and [Mg/Fe] (shaded panels). The black solid line shows predictions from our simple application of empirically calibrated Milky Way yields (Section~\ref{sec:mod_app_to_obs}), and the black dashed line shows predictions from modeling the full chemical evolution history  (Section~\ref{sec:metev}). Colored lines show predictions from \textsc{VICE} with theoretical yields for massive stars from \citet{nomoto_nucleosynthesis_2013}, \citet{chieffi13}, and \citet{Sukhbold2016}, and AGB and SNe~Ia yields from \citet{cristallo_evolution_2011,cristallo_evolution_2015} and \citet{Seitenzahl2013}. Where a colored line is absent, the corresponding yield table does not include that element; for Ti specifically, the \citet{chieffi13} predictions fall well below the plot limits. The theoretical yields fail to reproduce the observed abundances and are mutually inconsistent, in some cases differing by up to 0.8~dex. By contrast, the empirical yields reproduce the observed abundances for most elements, and incorporating metallicity evolution does not meaningfully change the predictions. Residuals for N, Sr, and Ba likely reflect AGB enrichment not captured by the two-channel model.}
    \label{fig:yields_results}
\end{figure*}

\begin{figure*}
    \centering
    \includegraphics[width=1\textwidth]{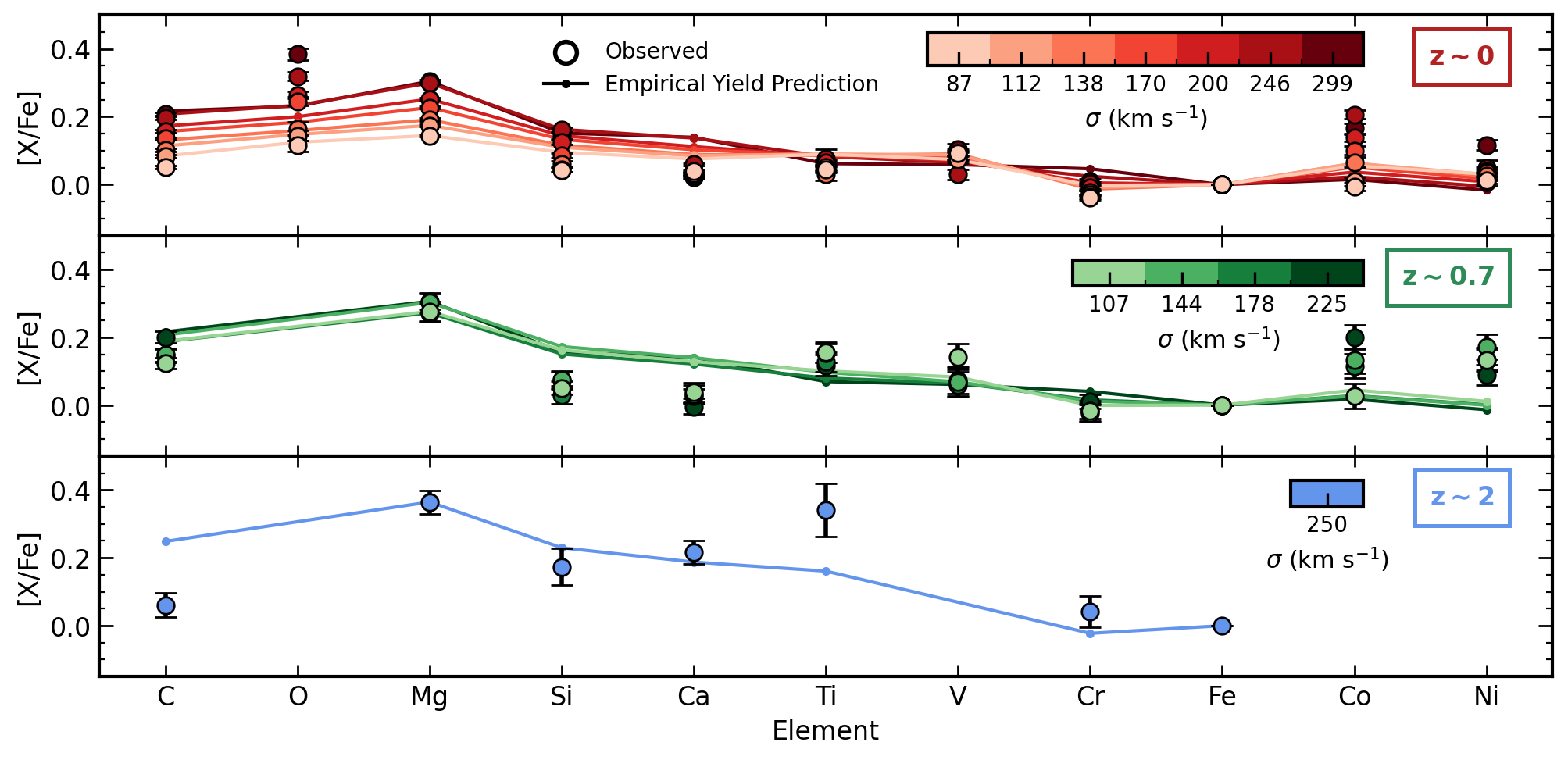}
    \caption{Observed (circles) and predicted (lines) abundance patterns for the SDSS ($z\sim0$; top), LEGA-C ($z\sim0.7$; middle), and SUSPENSE ($z\sim2$; bottom) samples. Predictions are based on empirically calibrated Milky Way yields, and assume no metallicity evolution. Overall, the empirical yields reproduce the observed abundance patterns well. We exclude elements with significant AGB contribution (N, Ba, Sr).
        }
    \label{fig:comparison}
\end{figure*}

The key idea is to treat each galaxy as we would an individual Milky Way star: an object whose abundance pattern reflects some mixture of CCSNe and SNe~Ia enrichment. Given the observed $[\mathrm{Mg/H}]$ and $[\mathrm{Mg/Fe}]$ of each galaxy, we solve for the relative contributions of CCSNe and SNe~Ia, $A_{\rm cc}$ and $A_{\rm Ia}$, via Eqs.~\ref{eq:acc} and \ref{eq:aia}, and then predict the full abundance pattern using the metallicity-dependent $\qxcc$ and $\qxia$ and Eq.~\ref{eq:xh}. This procedure is mathematically equivalent to interpolating, in linear abundances, between the high- and low-$\alpha$ Milky Way sequences to match the observed [Mg/Fe]. Indeed, the right panel shows that the predicted abundance pattern of the SDSS galaxies at [Mg/H]$=0$ falls between that of the high- and low-$\alpha$ Milky Way stars at the same metallicity (large red and blue circles) in nearly every element. We present results for the full galaxy samples in Figures~\ref{fig:yields_results}--\ref{fig:deltas}.

The advantage of this approach is that it is entirely empirical, requiring no chemical-evolution model---no assumed SFH or SNe~Ia delay-time distribution. However, a key limitation is that it assumes all stars formed from gas with a single chemical composition and therefore a single metallicity. In reality, galaxies undergo continuous enrichment, and their metallicity increases over time. Because some elements have strongly metallicity-dependent yields (Figure~\ref{fig:qs}), this assumption can bias the predicted abundance patterns. To account for the fact that galaxies are composed of stars with an intrinsic metallicity distribution, we introduce our second prediction, which incorporates a simple metallicity-evolution history. Comparing the two predictions allows us to assess how metallicity evolution influences the inferred abundance patterns.

\subsection{Incorporating metallicity evolution}
\label{sec:metev}

Here, we describe our method for predicting the multi-element abundances as a function of time, using a more realistic metallicity evolution history. Our approach is summarized in Figure~\ref{fig:model}. 
% In brief, we adopt a one-zone chemical evolution model that independently tracks Mg and Fe. We find a chemical evolution model that, after 14 Gyr, can reproduce our observed [Mg/H], [Mg/Fe], and stellar ages. Using this best-fit model, we then apply the metallicity-dependent yields to predict the time evolution of the full abundance pattern. 
We adopt a one-zone analytical chemical evolution model described by \citet{WAF2017}. In this model, Mg and Fe are tracked independently; CCSNe enrich the ISM with Mg and Fe instantaneously, while SNe Ia inject the ISM with Fe following a delay time distribution \citep[DTD,][]{maoz_supernova_2010} with a minimum delay time of $t_{\rm d,min}=0.15$~Gyr. We implement the model using the same setup as \citet{gountanis2025} and using the Python package \texttt{fanCE}\footnote{\url{https://github.com/nmgountanis/fanCE}} \citep{gountanis_modeling_2024}. We use the absolute Mg and Fe yields from Eq.~\ref{eqn:yvals}, and adopt the ``rise–fall'' star-formation history (SFH) introduced by \citet{Johnson21_vice},
\begin{equation}
\mathrm{SFH} \equiv \dot{M}_\star(t) \propto (1 - e^{-t/\tau_1})\; e^{-t/\tau_2},
\label{eqn:SFH}
\end{equation}
where $\tau_1$ and $\tau_2$ describe the characteristic rise and decline timescales of star formation. Most parameters are fixed to the \texttt{fanCE} defaults, but we allow the following to vary: the onset time of star formation ($t_{\rm start}$), the star-formation timescales ($\tau_1$ and $\tau_2$), the star-formation efficiency (SFE), and the outflow mass-loading factor ($\eta$).

%We empirically anchor our CCSN Mg and Fe yields using Milky Way observations following \citet{gountanis2025}. The Fe yield is adopted from \citet{Rodriguez23_SNe}, and the Mg yield is calibrated using the [Mg/Fe] ratio of very metal-poor halo stars from the H3 Survey, where Type Ia contributions are negligible ([Mg/Fe]${\rm cc}=0.55$). However, \citet{gountanis2025} showed that [Mg/Fe]${\rm cc}=0.55$ is insufficient to reproduce the high [Mg/H] values of nearby quiescent galaxies. For the present work, we therefore adopt a slightly higher value, [Mg/Fe]$_{\rm cc}=0.65$, to enable a plausible metallicity evolution history. \textbf{I should also see what abundance pattern I get using the burst model!}

Unlike in the Milky Way, where we can trace the full distribution of stars across different enrichment stages, for external galaxies we have only a single integrated measurement that represents the luminosity-weighted average of all stars. To enable a consistent comparison, we therefore compute the luminosity-weighted average of all stars in the model using the relation \citep[see also][]{gountanis2025}:
\begin{equation}
w(t)_{\mathrm{light}} = L_\mathrm{V}/M = 1.5\left(\frac{t_{\mathrm{obs}}-t}{\mathrm{Gyr}} + 0.1\right)^{-\beta} \frac{L_{\odot}}{M_{\odot}},
\label{eqn:weighing}
\end{equation}
with $\beta=0.90$. Here, $t_{\mathrm{obs}}$ is the age of the universe at the galaxy's observed redshift (i.e. $t_{\mathrm{obs}}=14$ for $z\sim0$), and $t_{\mathrm{obs}}-t$ is the age of the stellar population formed at time $t$.

With the chemical-evolution model in place, we construct metallicity-evolution histories that reproduce the observed [Mg/H], [Mg/Fe], and stellar age of each galaxy. This history is described by five parameters: the SFH parameters $\tau_1$, $\tau_2$, and $t_\mathrm{start}$, along with the star-formation efficiency $\tau_*$ and outflow mass-loading factor $\eta$. We explore this parameter space using the \texttt{emcee} MCMC sampler, adopting a Gaussian likelihood based on $\chi^2$. We adopt top-hat priors for all parameters.

\citet{gountanis2025} adopted $t_{\rm start} = 0.5$~Gyr and $[\mathrm{Mg/Fe}]_{\rm cc} = 0.45$ or $0.55$ as default values, and showed that it is challenging to reproduce the high $[\mathrm{Mg/Fe}]$ values of the most massive quiescent galaxies given their observed stellar population ages. We therefore adopt a higher $[\mathrm{Mg/Fe}]_{\rm cc}=0.65$ and leave $t_{\rm start}$ as a free parameter to ensure our models reproduce the observed $[\mathrm{Mg/H}]$ and $[\mathrm{Mg/Fe}]$ of our galaxy samples by construction. We note that our predicted abundances do not depend on these choices, as discussed in Section~\ref{sec:yields}.

Because the model has five free parameters but only three observables, it is formally underconstrained, and the posteriors are highly degenerate: many parameter combinations reproduce the observed [Mg/Fe], [Mg/H], and age equally well. However, these equally well-fit solutions produce nearly identical metallicity histories. We therefore treat the MCMC results as representative enrichment tracks rather than unique physical solutions, and use them to evaluate the metallicity-dependent yields.

Figure~\ref{fig:model} illustrates our approach and shows example results. The top row corresponds to the SDSS galaxy sample with the highest velocity dispersion, and the bottom row to the one with the lowest. The left-hand panels show the best-fit SFHs, while the middle panels show the corresponding evolution of light-weighted Mg and Fe abundances. The observed [Mg/H] and [Mg/Fe] of the seven SDSS galaxy bins are plotted as stars, with their corresponding chemical evolution tracks shown as lines. If we could resolve individual stars in these systems, their stellar populations would lie along these tracks. As expected, the high-velocity-dispersion galaxy has a shorter SFH, consistent with its higher [Mg/Fe]. In the SFH panels, we also plot the cumulative number of SNe Ia, computed by convolving the SNe DTD with the SFH. Galaxies with more extended star formation histories experience a larger cumulative SNe Ia contribution prior to quenching, meaning SNe Ia iron enrichment plays a more significant role in shaping their stellar abundance patterns.

With these best-fit metallicity evolution histories in hand, we can now predict the abundances of the remaining elements. From the best-fit [Mg/Fe] vs [Mg/H] tracks (middle panels of Figure~\ref{fig:model}), we determine the relative contributions of CCSNe and SNe Ia ($A_{\rm cc}$ and $A_{\rm Ia}$) at each timestep. We then use the metallicity-dependent yields inferred from $q_X^{\rm cc}$ and $q_X^{\rm Ia}$ via Eqs.~\ref{eq:yxcc} and \ref{eq:yxia} to predict the abundance pattern at each timestep. Finally, we compute the luminosity-weighted abundances using Eq.~\ref{eqn:weighing}, which represent the observable, integrated stellar composition predicted by the model. The resulting evolution of the light-weighted abundance patterns is shown in the right-hand panels of Figure~\ref{fig:model}. As expected, the galaxy with the more extended SFH evolves to lower [$\alpha$/Fe], reflecting the greater cumulative contribution from SNe~Ia. We also compare these predictions to the approximation described in Section~\ref{sec:mod_app_to_obs}, which estimates the multi-element abundance pattern directly from the observed [Mg/H] and [Mg/Fe] without modeling the enrichment history. The overall close agreement between the two approaches for the $\alpha$ and Fe-peak elements validates the use of the simpler method. We discuss the comparison of these approaches in more detail in Section~\ref{sec:z0}.

\subsection{Theoretical yield predictions}
\label{sec:vice}

In addition to the empirical yield predictions, we also test theoretical yield tables, which are more commonly used in chemical evolution modeling. To compute the theoretical yield predictions, we use \textsc{VICE} \citep{johnson2020}, a flexible chemical evolution code that separately tracks enrichment from CCSNe, SNe Ia, and AGB stars. We model each galaxy as a one-zone system and adopt the same SFH parameterization as described above, treating the SFH, SFE, and outflow mass-loading factor as free parameters. The SNIa and AGB yields are fixed to \citet{Seitenzahl2013} and \citet{cristallo_evolution_2011,cristallo_evolution_2015}, respectively, and we test three CCSN yield sets: \citet{nomoto_nucleosynthesis_2013}, \citet{chieffi13}, and \citet{Sukhbold2016}. For each set of CCSN yields, we use MCMC sampling to determine the light-weighted chemical evolution track that best reproduces the observed light-weighted Mg and Fe abundances, following the same fitting procedure as in the previous section.

A practical complication arises for the \citet{chieffi13} and \citet{Sukhbold2016} yield tables: both fail to reproduce the observed Mg abundances of the SDSS galaxies. This Mg underproduction problem is well documented in the Milky Way, where theoretical yield tables produce $[\mathrm{O/Mg}]$ ratios that are too high by factors of $2.5$--$4$ \citep{griffith2021}, a discrepancy that persists across explosion landscapes and cannot be resolved by varying the IMF or black hole formation prescription. For these two yield sets, we therefore apply an ad hoc correction, rescaling the IMF-averaged Mg yield to satisfy $[\mathrm{Mg/Fe}]_{\rm cc} = [\mathrm{O/Fe}]_{\rm cc}$:
\begin{equation}
    y_{\rm Mg} = y_{\rm O} \times \frac{Z_{\rm Mg,\odot}}{Z_{\rm O,\odot}},
\end{equation}
where $y_{\rm O}$ is the IMF-averaged O yield from the same table and $Z_{\rm Mg,\odot}$ and $Z_{\rm O,\odot}$ are the solar mass fractions. This is motivated by the fact that both O and Mg are primary $\alpha$ elements synthesized in massive stars on similar timescales. We do not apply this correction to the \citet{nomoto_nucleosynthesis_2013} yields, which include a hypernova contribution that produces sufficient Mg without adjustment. The need for this correction underscores the limitations of current theoretical yield tables, particularly for massive quiescent galaxies where Mg is often the only measurable $\alpha$ abundance. We present these results in Figure~\ref{fig:yields_results}.

\newpage
\section{Results}
\label{sec:results}

\begin{figure*}
    \centering
    \includegraphics[width=\textwidth]{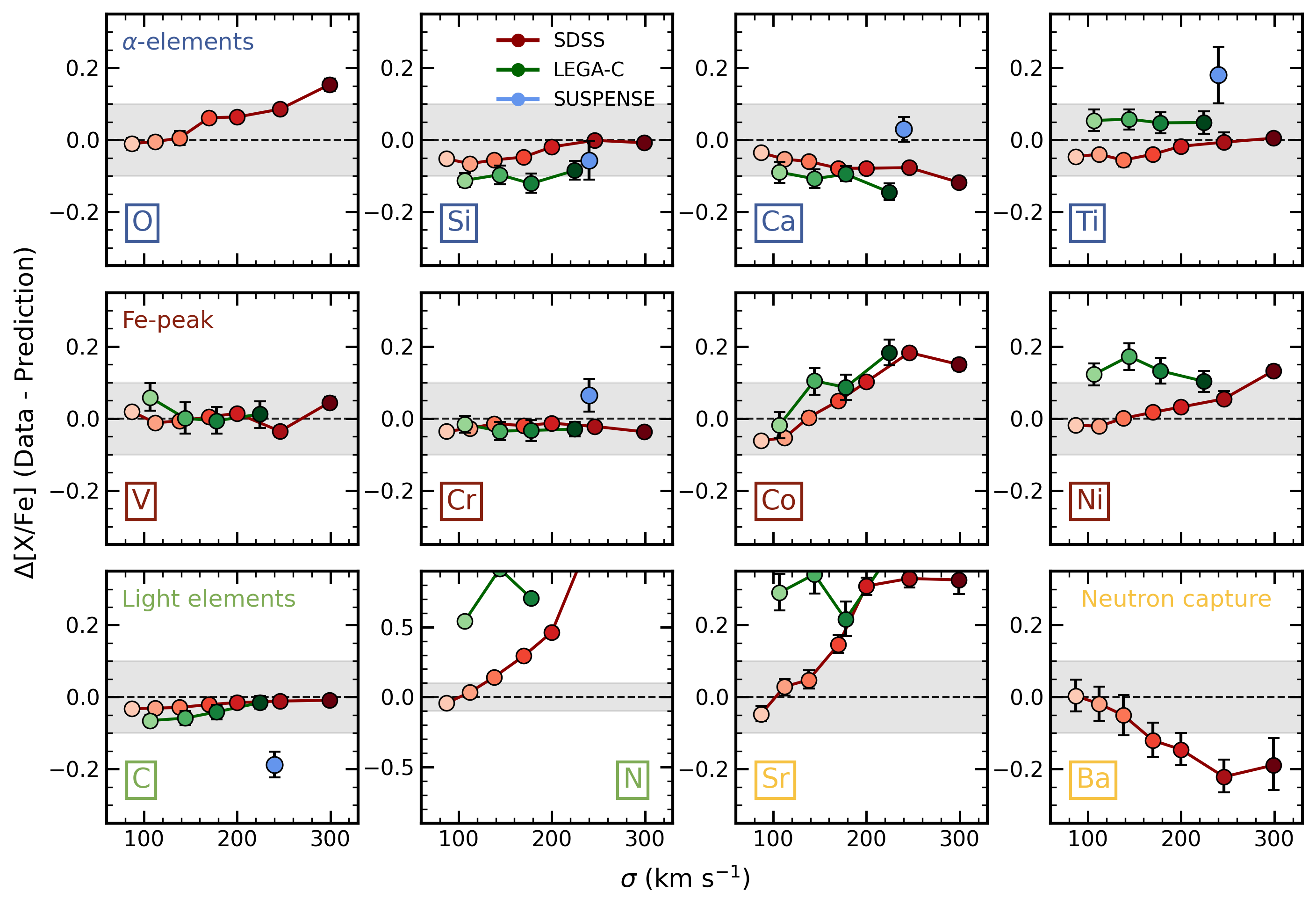}
    \caption{Residuals between observed and predicted abundance ratios, $\Delta$[X/Fe] = (data – model), corresponding to Figure~\ref{fig:comparison}. Lines and points represent the SDSS ($z\sim0$; red), LEGA-C ($z\sim0.7$; green), and SUSPENSE ($z\sim2$; blue) samples. Predictions are based on the empirically calibrated Milky Way yields assuming no metallicity evolution. The dashed line marks zero residual, and a positive residual implies that the model underpredicts the observed abundance. A region corresponding to 0.1\;dex agreement is highlighted in gray. Overall, most elements are reproduced to within 0.1\;dex, with N, Sr, and Ba showing significant offsets, especially at high velocity dispersion.}
    \label{fig:deltas}
\end{figure*}

In this Section, we compare predicted and measured multi-element abundance patterns for the SDSS ($z\sim0$), LEGA-C ($z\sim0.7$), and SUSPENSE ($z\sim2$) samples. In Figure~\ref{fig:yields_results}, we focus on $z=0$ and compare empirical and theoretical yield predictions against the data. In Figure~\ref{fig:comparison}, we extend the empirical yield predictions to all three redshifts, and in Figure~\ref{fig:deltas} we show the corresponding residuals as a function of velocity dispersion. Finally, we examine the impact of AGB enrichment on elements not well described by the two-channel framework.

\subsection{Comparison at $z=0$}
\label{sec:z0}

Figure~\ref{fig:yields_results} compares observed multi-element abundance patterns from the SDSS sample with predictions from empirical and theoretical yield prescriptions. Each panel shows a single element grouped by its dominant nucleosynthetic origin. The black lines show predictions based on the empirically calibrated Milky Way yields introduced in Section~\ref{sec:yields}: black solid lines represent the simplified method (Section~\ref{sec:mod_app_to_obs}), which applies these yields at the observed [Mg/H] and [Mg/Fe], while the black dashed lines represent complete evolutionary models that track the full metallicity history (Section~\ref{sec:metev}). Colored lines show \textsc{VICE} predictions using theoretical stellar yields (Section~\ref{sec:vice}). Where a line is absent, the corresponding yield table does not include that element, with the exception of Ti, where the \citet{chieffi13} predictions fall well below the plot limits. In all cases, predictions are anchored to the observed [Mg/H] and [Mg/Fe] (shaded panels), and therefore pass through those points by construction. As we show below, the empirically calibrated yields reproduce the observed abundance ratios for most elements, while theoretical yield predictions are mutually inconsistent and fail to match the data.

We first compare the two applications of the empirical yields. If nucleosynthetic yields were independent of metallicity, the two approaches would be identical; the degree to which they differ, therefore, tests whether metallicity evolution matters in practice. For most elements, the solid and dashed black lines agree to within $\sim$0.1~dex, indicating that metallicity-dependent yields have little impact on the predicted abundances. The two approaches diverge most significantly for Ba and N, which have the strongest metallicity dependence in the empirically inferred yields (Figure~\ref{fig:qs}). Surprisingly, where the two methods differ, the simplified model generally achieves equal or better agreement with the observations, with Co being the notable exception. This could indicate that our metallicity-evolution models are not sufficiently accurate for strongly metallicity-dependent yields, or it could reflect systematics in the APOGEE abundance measurements, or genuine differences in yields between Milky Way disk stars and the massive quiescent galaxy population. Given that the full evolutionary models do not improve the predictions, we focus on the simplified single-metallicity approach for the remainder of this section.

For most $\alpha$-elements and Fe-peak elements, we find good agreement between the empirical yield predictions and the observed abundance ratios. For galaxies near solar metallicity, the agreement is to within 0.05 dex for all of these elements. For the most massive galaxies, which occupy a region of [Mg/Fe]-[Mg/H] well outside the range of Milky Way disk stars (see Figure~\ref{fig:apogee_vs_galaxies}), we find differences as large as 0.1-0.2 dex for O, Ca, Co, and Ni. The empirical yield models fail significantly at high galaxy masses for three elements: Sr, Ba, and N.  All of these are elements where we expect the delayed contribution to come from AGB stars rather than SNe Ia. We address the case of N in Section~\ref{sec:agb}, where we show that explicitly modeling AGB enrichment provides much better agreement with the data.

The \textsc{VICE} models based on theoretically predicted yields reproduce the observed abundance ratios significantly less well than the empirically calibrated models. The spread among predictions from the three theoretical yield sets -- which differ by up to 0.8~dex in some cases -- itself illustrates the level of uncertainty in current nucleosynthesis calculations, and it is therefore unsurprising that these models typically struggle to match the data at the 0.2--0.4~dex level. This failure is not unexpected, given the known sensitivity of nucleosynthetic yields to uncertain input physics. We conclude that theoretically predicted yields are not yet accurate enough to interpret the abundance ratios of massive quiescent galaxies, but empirically calibrated yields based on Milky Way stars finally provide a useful basis for doing so.

%The solid black line applies these yields at a single metallicity (the observed value of [Mg/H] and [Mg/Fe]; Section~\ref{sec:mod_app_to_obs}), while the dashed line incorporates the full metallicity evolution history (Section~\ref{sec:metev}). The theoretical yield predictions (Section~\ref{sec:chempy}), which were modeled using \textsc{Chempy} fits to the observed [Mg/H] and [Mg/Fe], are also shown (blue and green dashed lines). 

%Overall, the empirical yields provide a substantially better match to the observed abundance patterns than the theoretical yield sets. For the $\alpha$ and Fe-peak elements, the median absolute offset between the empirical predictions and observations is only 0.04~dex, compared to 0.18~dex for theoretical yields. This improvement highlights that the abundance patterns of Milky Way stars capture the integrated enrichment of massive quiescent galaxies far more effectively than current theoretical yields. 

%Somewhat counterintuitively, including a full metallicity-evolution history (dashed black line) does not improve agreement; for nearly all elements (except Co and N), it performs similarly to, or slightly worse than, assuming a single metallicity (solid black line). The elements for which the two approaches diverge (C, N, Co, V, Sr, Ba) are those with strongly metallicity-dependent yields (refer to Figure~\ref{fig:qs}). 
% In other words, Mg and Fe already encode the key information needed to predict the other $\alpha$ and Fe-peak elements. 

\begin{figure*}
    \centering
    \includegraphics[width=0.8\textwidth]{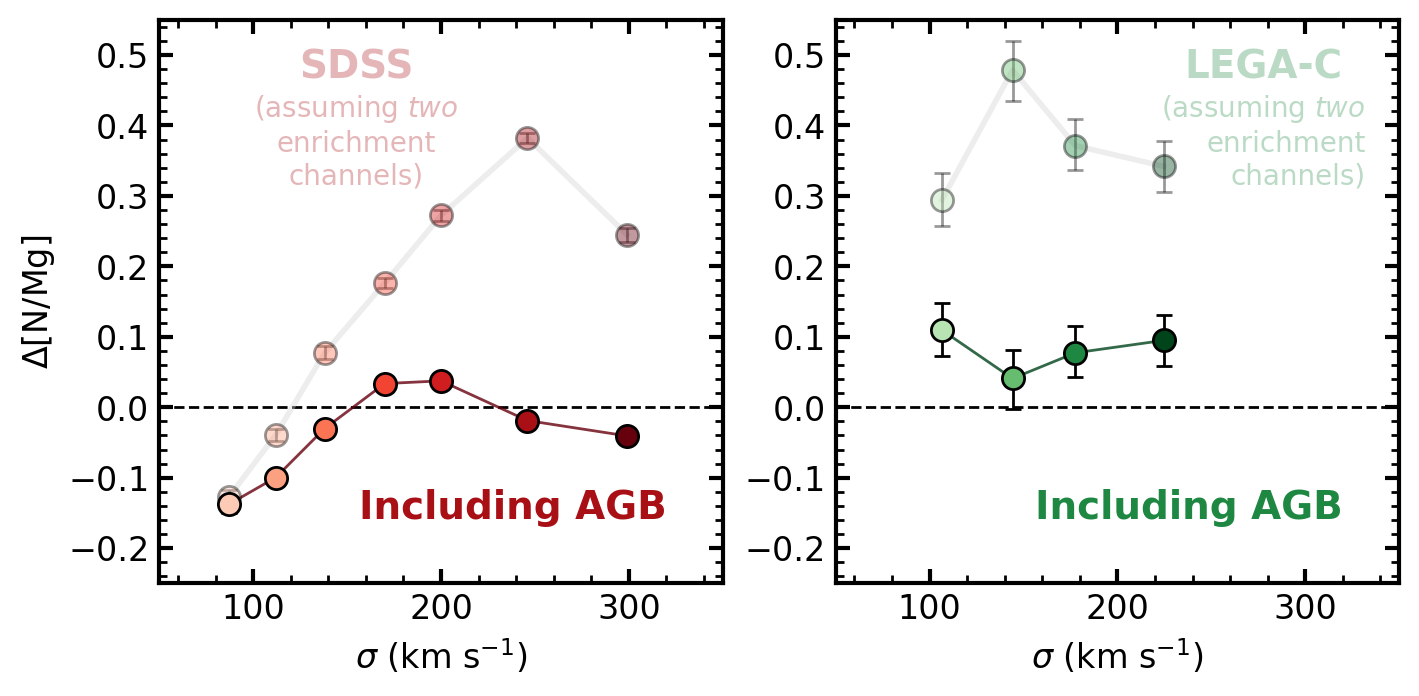}
    \caption{Residuals in $[\mathrm{N/Mg}]$ for SDSS (left, red) and LEGA-C (right, green) galaxies as a function of velocity dispersion. Faint lines show the residuals from the empirical two-channel model (Section~\ref{sec:metev}), which systematically underpredicts nitrogen at all velocity dispersions. Solid lines show results from a \textsc{VICE} model that explicitly includes AGB enrichment, using the CCSN nitrogen yields of \citet{johnson_empirical_2023} and the metallicity-dependent AGB yields of \citet{cristallo_evolution_2011, cristallo_evolution_2015}, renormalized to reproduce solar $[\mathrm{N/Mg}]$ in a Milky Way-like SFH (see Section~\ref{sec:agb}). Including this third enrichment channel accounts for the observed nitrogen excess at both redshifts, including its dependence on velocity dispersion.}
    \label{fig:nmg}
\end{figure*}

%The success of the single-metallicity predictions stems from the assumptions underlying our yield calibration. Implicitly, we assume that a star's chemical composition reflects enrichment only by other stars at the \textit{same} metallicity. This assumption breaks down when the actual yield of an element varies strongly with metallicity. Most $\alpha$ and Fe-peak elements have mild metallicity-dependence, so including a full metallicity-evolution history gives a similar result to assuming a single metallicity. The light- and neutron-capture elements, however, exhibit stronger metallicity dependence (Figure~\ref{fig:qs}), leading to larger discrepancies between the two predictions. For this reason, we adopt the single-metallicity predictions for the remainder of the analysis.

\subsection{Comparisons across redshift}

Figure~\ref{fig:comparison} compares the observed (circles) and predicted (lines) multi-element abundance patterns for SDSS (red), LEGA-C (green), and SUSPENSE (blue). Again, these predictions are based on empirically calibrated Milky Way yields applied at a single metallicity, without modeling the galaxies’ chemical-evolution histories. In Figure~\ref{fig:deltas}, we show the residuals of these predictions as a function of velocity dispersion, with the gray shaded region indicating 0.1~dex agreement. Mg and Fe are excluded because the model is anchored to their observed values, resulting in zero residuals by construction.

Overall, predictions based on empirically calibrated Milky Way yields show excellent agreement with the observed chemical abundances at all three redshifts, with median absolute offsets of just 0.03~dex for SDSS, 0.06~dex for LEGA-C, and 0.06~dex for SUSPENSE across the $\alpha$- and Fe-peak elements. The slightly higher residuals at $z\sim2$ likely reflect the larger statistical uncertainties (see Figure~\ref{fig:data}). The previously noted discrepancies for N and Sr are also present in the comparison to LEGA-C. Like SDSS, LEGA-C shows a trend of increasing [Co/Fe] with increasing velocity dispersion that is not predicted by the models, even though LEGA-C shows little trend of [Mg/Fe] with velocity dispersion. The models also overpredict [C/Fe] in the SUSPENSE galaxies at $z\sim2$ by 0.2\;dex, suggesting differences in C enrichment at early times or in the conditions that give rise to quiescent galaxies at high redshift.

% It is not surprising that C, N, Sr, and Ba are the elements that are least well reproduced by the two-process prediction. A substantial fraction of these elements is synthesized in a third, intermediate enrichment channel, AGB stars. In Section~\ref{sec:agb}, we test whether introducing an explicit AGB contribution resolves the remaining discrepancies.

\subsection{The impact of AGB enrichment}
\label{sec:agb}
To understand the large residuals in the N, Sr, Ba, and (at $z\sim2$) C abundances, we must confront a major limitation of the empirical yields: that there are only two enrichment channels, a prompt channel (traced by Mg) and a delayed channel (traced by Fe). In reality, a third channel---AGB stars---plays a significant role in galactic chemical evolution and is not traced by either Mg or Fe. AGB stars enrich on intermediate timescales and are significant producers of several light and slow neutron-capture elements. It is no coincidence that these are precisely the elements where our predictions fail. 
%Model predictions for C agree well with the data at $z\sim0$ and $z\sim0.7$, but the discrepancy at $z\sim2$ for SUSPENSE could also reflect the importance of AGB enrichment. 

Here we examine whether a more physically accurate treatment of AGB enrichment can explain our models' underprediction of N. For this purpose, we again use VICE, which supports multiple enrichment channels. We adopt the chemical-evolution parameters ($\tau_1$, $\tau_2$, $t_\mathrm{start}$, $\tau_*$, and $\eta$) inferred for each velocity-dispersion bin in Section~\ref{sec:metev} and use them as direct inputs to VICE.

Nitrogen is produced only by CCSNe and AGB stars. For the CCSN yields, we adopt the empirically calibrated N yields from \citet{johnson_empirical_2023}, $\yncc=3.6\times 10^{-4}$, independent of metallicity. For the AGB N yields, we adopt metallicity-dependent theoretical yields from \citet{cristallo_evolution_2011,cristallo_evolution_2015}. 

Before applying these yields to our quiescent-galaxy models, we verify that their normalization is physically reasonable. In particular, a Milky Way–like disk SFH should reproduce solar [N/Mg] at [Mg/H]$\;=0$. Running a disk-like SFH in VICE with the above yields, we find that [Mg/Fe] approaches 0 at [Mg/H]$\;=0$ as desired, but [N/Mg] is offset by 0.2\;dex. We therefore renormalize the CCSN and AGB nitrogen yields by $-0.2$\;dex so that a disk-like SFH yields [N/Mg]$\;=0$ at [Mg/H]$\;=0$.\footnote{The difference from the calibration of \citet{johnson_empirical_2023} reflects our different choice of the Mg yield $\ymgcc$.}

With the nitrogen yields calibrated, we compute the resulting VICE predictions for [N/Mg] for the SDSS and LEGA-C samples. Figure~\ref{fig:nmg} shows the residuals between the data and these VICE predictions (solid, high-opacity lines), and for comparison, we overplot the original residuals from Figure~\ref{fig:deltas} (faint, transparent lines). Remarkably, once we include the AGB component, the excess nitrogen is accounted for at both redshifts, including its dependence on velocity dispersion.

Why do the empirical yields fail so dramatically at predicting nitrogen abundances in quiescent galaxies, yet perform so well in the Milky Way? The key is the strong metallicity dependence of AGB N yields. At low metallicity, AGB stars produce almost no N. In the Milky Way, the ISM does not reach the metallicities needed for efficient AGB N production until after SNe Ia begin enriching the gas. Consequently, N \textit{appears} to track the delayed (SNIa–like) channel, and thus the empirical yields lump AGB-produced N into the SNe Ia component \citep[][]{johnson_empirical_2023}. In massive quiescent galaxies, this is not the case. Their ISM reaches super-solar metallicity rapidly, allowing AGB stars to contribute substantial N early. In such systems, N should instead behave more like a prompt-channel element (CCSN-like). Thus, applying MW-calibrated yields---which assign most of the AGB-produced N to the delayed channel---to quiescent galaxies, which experience minimal delayed enrichment, guarantees that we will underpredict their true N enrichment. This effect is amplified in the most rapidly forming galaxies, leading to the strong velocity dispersion dependence.

%A similar argument can be used to explain the observed carbon deficiency at $z\sim2$. In the Milky Way, AGB stars efficiently produce C even at low metallicity, so their contribution is effectively folded into the prompt yield \citep[see][]{boyea25}. In rapidly forming $z\sim2$ galaxies, however, AGB stars have contributed only minimally to the chemical enrichment of the stellar population, causing predictions calibrated on Milky Way stars to overestimate their true carbon abundance. At lower redshift, where star formation is more extended, and AGB enrichment has had more time to accumulate, this approximation is more appropriate. 

Similar considerations may explain the discrepancies between our models and the observed trends for Sr and Ba in SDSS and LEGA-C, and the discrepancy with the observed C abundance in SUSPENSE.  In each of these cases, the delayed contribution to element production is expected to come from AGB stars rather than SNe Ia, with a characteristic delay time that is intermediate between that of CCSN and SNe Ia.\footnote{Comparisons of theoretically predicted DTDs can be found in Fig.~5 of \cite{johnson2020} for Sr and Fig.~3 of \cite{boyea25} for C.}  As illustrated above for N, the impact of AGB enrichment on our predictions can involve a complicated interplay between time delays and metallicity-dependent yields, which affect both the two-process calibration of yields in the Milky Way and the evolution in the more rapidly enriching SFH of quiescent galaxies.  It is somewhat surprising that the discrepancies we find for Sr and Ba are opposite in sign, since the two elements have qualitatively similar metallicity dependence in their $\qxcc$ and $\qxia$ (see Figure~\ref{fig:qs}).  However, the inferred metallicity dependence is strong and non-monotonic, in agreement with theoretical AGB yield models (see, e.g., \citealt{griffith19_galah}), which may make predictions sensitive to this metallicity dependence and to the relative contributions of prompt and delayed enrichment.  

Our results for other elements highlight the importance of empirically calibrated yields for chemical evolution models, but for these elements, the calibration will need to be based on Milky Way chemical evolution models rather than a two-process decomposition alone, as done for N by \cite{johnson_empirical_2023} and for C by \cite{boyea25}.  We leave the investigation of the evolution of C, Sr, and Ba for future work, building on the example of N explored here.

% Interestingly, we find that the Ba residuals go in the opposite direction of N and Sr: at higher velocity dispersion, we \textit{over}-predict the amount of Ba. Again, this can be explained by the fact that Ba AGB yield generally \textit{falls} with increasing metallicity, whereas N AGB yields rise.

\section{Discussion}
\label{sec:discussion}

In this paper, we introduced a framework for predicting the multi-element abundance patterns of quiescent galaxies using empirically calibrated nucleosynthetic yields derived from Milky Way stars. Rather than relying on uncertain theoretical yields or explicit chemical-evolution modeling, we treat each galaxy as a stellar population whose abundances reflect a mixture of CCSNe and SNe~Ia enrichment, using the observed [Mg/H] and [Mg/Fe] to anchor the prediction. We find that this framework reproduces the $\alpha$ and Fe-peak elements remarkably well across three redshifts --- SDSS ($z\sim0$), LEGA-C ($z\sim0.7$), and SUSPENSE ($z\sim2$) --- with a median absolute offset of only 0.03--0.06~dex, outperforming all commonly used theoretical yield sets; including a full chemical-evolution history does not meaningfully improve the predictions. Despite this overall success, several elements show significant residuals, most notably N, Sr, Ba, and (at $z\sim2$ only) C. We show that these failures are not random: they likely arise because the two-process framework does not separately account for AGB enrichment. In the case of N, we show that the magnitude and sign of the residuals can be understood through the strong metallicity dependence of N yields.  In this Section, we place our findings in a broader context.

\begin{figure*}
    \centering
    \includegraphics[width=1\textwidth]{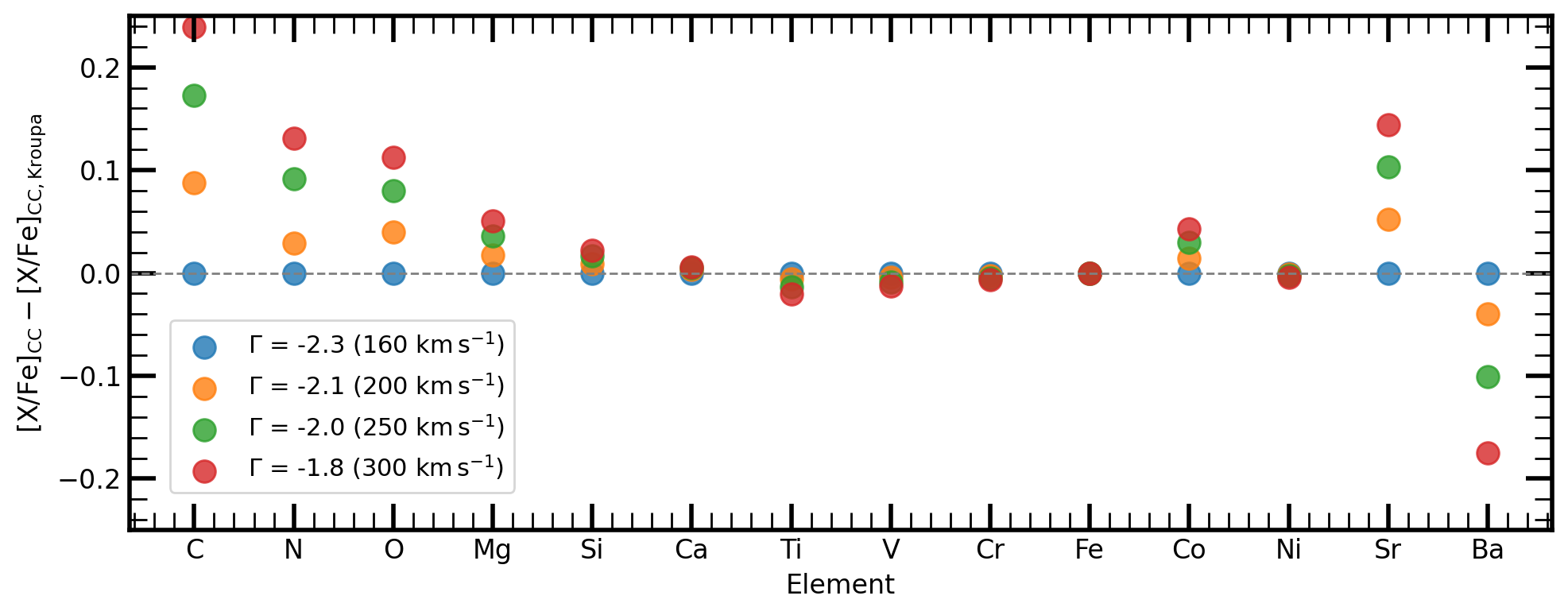}
    \caption{Change in IMF-averaged CCSN yield ratios $[\mathrm{X/Fe}]_{\rm cc}$ relative to a Kroupa IMF ($\Gamma = -2.3$), computed using the \citet{Sukhbold2016} yields under the N20 explodability prescription. Each colored point corresponds to a different high-mass IMF slope, with associated velocity dispersions following the \citet{vanDokkum2024} parameterization. The dashed line marks zero, i.e., no change relative to Kroupa. Lighter elements (C, N, O) show the largest enhancements with increasing top-heaviness, while most Fe-peak elements are largely unaffected. Sr and Ba respond in opposite directions, and Co shows a mild enhancement.}
    \label{fig:imf}
\end{figure*}

\subsection{Variations in the high-mass IMF}
\label{sec:imf}

Recent \textit{JWST} observations have revealed several surprising features of early galaxy populations, including a larger-than-expected number of luminous galaxies at $z>9$ \citep[e.g.,][]{Harikane2023,Mcleod2024,finkelstein2024}, unusually blue, UV-bright systems at $z>9$ (``blue monsters’’; \citealt{Ziparo2023,Ferrara2025}), and the presence of massive galaxies at early times that appear to require impossibly high star-formation efficiencies \citep[e.g.,][]{glazebrook_massive_2024,carnall_jwst_2024}. One natural explanation is an excess of massive stars in these galaxies \citep[e.g.,][]{denbrok_recovery_2024, trinca2024, Hutter2025}. \citet{vanDokkum2024} further propose that the high-mass (and low-mass) IMF slope varies with galaxy velocity dispersion, such that higher-dispersion galaxies have shallower high-mass IMF slopes. Independent support comes from measurements of $^{13}$CO/C$^{18}$O isotopologue ratios in dusty starburst galaxies at $z\sim2-3$ \citep{zhang18, guo24}, the likely progenitors of the massive quiescent systems studied here, which also imply a top-heavy IMF.

Variations in the high-mass IMF have direct consequences for nucleosynthesis: because CCSN yields depend sensitively on progenitor mass \citep[e.g.,][]{nomoto_nucleosynthesis_2006}, any shift toward more or fewer high-mass stars alters the IMF-averaged yield ratios of CCSNe. Our empirical abundance predictions implicitly assume a Milky Way–like IMF, because they are calibrated \textit{in} the Milky Way. If massive quiescent galaxies instead formed with a systematically different IMF, this would manifest as residuals relative to our Milky Way-based predictions. While a more top-heavy IMF increases the absolute yields of all elements, only changes in \textit{yield ratios} matter here. Because all abundances are normalized to the total metal enrichment traced by $[\mathrm{Mg/H}]$ (Eq.~\ref{eq:acc}), a uniform rescaling of overall metal production is absorbed by this normalization, and only shifts in $[\mathrm{X/Fe}]_{\rm cc}$ produce observable residuals.
 
To assess how IMF variations affect our predictions, we compute IMF-averaged CCSN yield ratios [X/Fe]$_{\rm cc}$ as a function of the high-mass slope. We adopt the \citet{vanDokkum2024} IMF, in which higher-dispersion galaxies have shallower high-mass slopes (i.e., more top-heavy IMFs). 
For the yields, we use \citet{Sukhbold2016} under their N20 explodability prescription and compute IMF-averaged element production using \texttt{VICE}. The \citet{Sukhbold2016} yields are available only at solar metallicity, but their fine sampling of progenitor masses makes them the best grid for studying IMF effects. The resulting yield ratios are shown in Figure~\ref{fig:imf}, with the corresponding IMFs illustrated in the inset; the dashed line denotes the fiducial Kroupa case.

As the high-mass slope becomes shallower, the relative production of lighter elements (C, N, O, Mg, Si) increases, while that of iron-peak elements (V, Cr, Ni) remains largely unchanged. This reflects the growing importance of the most massive stars ($M_* \gtrsim 20\,M_\odot$), which preferentially synthesize $\alpha$ elements relative to lower-mass CCSN progenitors ($8 \lesssim M_*/M_\odot \lesssim 20$). In the case of carbon and nitrogen, much of the production in the \citet{Sukhbold2016} models is released in massive star winds, even if the stars eventually collapse to black holes rather than exploding as CCSN (see Fig. 2 of \citealt{griffith2021}). The mass loss prescriptions for massive stars are uncertain, so while the changes to [X/Fe]$_{\rm cc}$ are significant for those elements, they should be interpreted with caution. 

The behavior of Sr and Ba is particularly striking. A more top-heavy IMF suppresses Ba while enhancing Sr, which qualitatively matches the opposite-sign residuals in Figure~\ref{fig:deltas}. If the IMF becomes more top-heavy in higher-dispersion galaxies, as proposed by \citet{vanDokkum2024}, this would naturally explain why the Sr and Ba residuals grow with velocity dispersion.

Among the Fe-peak elements, Co stands out in Figure~\ref{fig:imf}. Unlike V, Cr, and Ni, [Co/Fe] shows a mild but non-negligible IMF dependence. At the same time, Co exhibits the largest residuals among the Fe-peak elements in Figure~\ref{fig:deltas}. The strong [Co/Fe] trend with velocity dispersion in $z\sim0$ massive quiescent galaxies has been puzzling since it was first reported by \citet{conroy_early-type_2014}, and it remains difficult to explain \citep[][]{denbrok_recovery_2024}. A more top-heavy IMF that varies with velocity dispersion could account for some of the [Co/Fe] residuals as a function of velocity dispersion, though the effect is not quite large enough to fully explain the observed trend.

Beyond the Fe-peak elements, a more top-heavy IMF also boosts the O yield, consistent with the observed trend of increasing [O/Fe] residuals with velocity dispersion in Figure~\ref{fig:deltas}. This suggests that the IMF variation proposed by \citet{vanDokkum2024} could also explain this trend. However, the residuals of C are small and only weakly dependent on velocity dispersion at $z\sim0$ and $z\sim0.7$, which is difficult to reconcile with this perspective if the \citet{Sukhbold2016} predictions are accurate. For N, the predicted trends are correct in sign but small in magnitude, and we have already argued that the discrepancies for N in Fig.~\ref{fig:deltas} are explained by AGB enrichment.

Several of the abundance residuals are qualitatively consistent with a more top-heavy IMF in higher-dispersion galaxies, as is proposed by \citet{vanDokkum2024}. However, theoretical CCSN yields remain highly uncertain, and a more quantitative analysis is required to test this interpretation. In the case of Sr and Ba, disentangling the IMF and AGB contributions will require modeling both effects simultaneously; theoretical yields suggest that the massive star contribution to these elements is subdominant but not negligible. Importantly, IMF variations also alter the overall metal yield per explosion, which affects the time evolution of chemical enrichment and the star-formation timescales inferred from [Mg/Fe]; we explore this effect in Beverage et al. (\textit{in prep}).

\subsection{Universality of yields}

It was not obvious a priori that the inferred nucleosynthetic yields from Milky Way stars would reproduce the observed element abundance patterns of massive quiescent galaxies. These extreme systems formed the bulk of their stars in short, intense bursts at early epochs, under gas pressures, densities, and metallicities very different from those of the Milky Way disk \citep[e.g.][]{thomas_environment_2010-1, mcdermid_atlas3d_2015, tacchella2017, carnall_jwst_2024}. For example, quiescent galaxies at $z\gtrsim2$ are found to have formed stars in just 0.1-0.5 Gyr \citep[][]{kriek_massive_2016, beverage_suspense_2025, hamadouche2026arXiv, leung2026arXiv}, conditions far more extreme than those of the Milky Way. Furthermore, recent observations of the earliest galaxies show extreme [N/O] enhancements (e.g., GNz-11; \citealt{bunker2023}) that are difficult to reconcile with standard stellar models, possibly pointing to non-universal yields in the earliest star-forming environments.

In principle, numerous factors could alter yields across such environments. Stellar wind mass-loss rates depend sensitively on metallicity and remain highly uncertain \citep[][]{Maeder1992,vink2022,josiek2024,huscher2025}. Rotation and binary interactions are likely enhanced in dense, early star-forming regions \citep[see][and references therein]{marchant2024}, both of which affect nucleosynthetic yields \citep[e.g.][]{Maeder2012, demink2013}. Uncertainties in the CCSN explosion landscape --- including fallback, progenitor compactness, and the ``islands of explodability'' --- add further complexity \citep[][]{sukhbold2014,Sukhbold2016,ertl2016,griffith2021}.

Despite these potential differences, our results show that Milky Way calibrated yields reproduce the $\alpha$ and Fe-peak abundance ratios of massive quiescent galaxies remarkably well. Strikingly, \citet{hasselquist24} find a similar result in Milky Way dwarf satellite galaxies: Milky Way calibrated yields reproduce the $\alpha$ and Fe-peak elements well, with residuals mainly found for elements with strongly metallicity-dependent yields, for which the two-process framework has known limitations.\footnote{\citet{hasselquist24} do find that the Milky Way calibrated two-process model overpredicts [Co/Mg] in the Sgr, GSE, and LMC dwarfs, by about 0.2\;dex, which continues the trend shown in our Fig.~\ref{fig:deltas}.} Taken together, results in the Milky Way disk \citep[][]{weinberg_chemical_2019, weinberg22_apogee, sit25}, dwarf satellites \citep[][]{hasselquist24}, and massive quiescent galaxies (this paper) suggest that the dominant nucleosynthetic channels shaping the $\alpha$ and Fe-peak elements are robust to a range of star-forming conditions. 

% Together, these results suggest that the IMF-averaged nucleosynthetic yields of CCSNe and SNe~Ia are remarkably universal across environments---from low-mass dwarf galaxies, to the Milky Way disk, to the most massive quiescent galaxies across cosmic time.

% This universality is further supported by the small amplitude of the IMF-driven abundance variations explored in Section~\ref{sec:imf}. The predicted shifts in [X/Fe]$_{\rm cc}$ from IMF variations like those advocated by \citet{vanDokkum2024} are only $\sim$0.05~dex---comparable to our observed residuals. Our results here provide no strong evidence for or against IMF variations at this level, though they limit more extreme IMF variations that could produce larger shifts of abundance ratios. 

% The absence of systematic residuals at this level in the $\alpha$ and Fe-peak elements therefore places empirical constraints on any IMF-driven yield variations. Taken together, these results suggest that the dominant nucleosynthetic channels shaping the $\alpha$ and Fe-peak elements are robust to the range of star-forming conditions.

\subsection{How do multi-element abundances inform SFHs?}

A surprising result of this work is how little information is needed to reproduce the $\alpha$ and Fe-peak abundance patterns of quiescent galaxies. Once Mg and Fe are known, the remaining elements follow almost automatically, and incorporating a full chemical-evolution history produces almost no improvement. This behavior arises because Mg and Fe already encode the relative contributions of CCSNe and SNe~Ia, and the yields of most $\alpha$ and Fe-peak elements have minimal metallicity dependence. The other elements do not provide independent information; they merely reinforce the same measurement.

This has a practical implication: the $\alpha$ and Fe-peak elements, taken together, constrain essentially one quantity --- the CCSN-to-SNe~Ia enrichment ratio. This ratio is a powerful diagnostic of star-formation timescales, hence the long-standing utility of [$\alpha$/Fe]. But we find that measuring a dozen elements rather than just Mg and Fe does not meaningfully sharpen that constraint; instead, it provides evidence for the universality of yields.  
% we need tracers that enrich at different timescales or whose yields are metallicity-dependent.

In principle, to recover finer structure in a galaxy's SFH, elements with strong metallicity-dependent yields could help: if yields varied sharply with metallicity, their abundances would carry information about the galaxy’s full enrichment history, not just its prompt-to-delayed ratio. However, the elements with the strongest metallicity dependence (e.g., Al and Na) are challenging to measure in quiescent galaxies at these redshifts and S/N levels. For the elements we can measure, the metallicity dependence is too weak to provide additional constraints on the SFH.

A more promising route to recovering finer SFH detail is to use elements that enrich on timescales different from those of CCSNe or SNeIa. This is where AGB stars will become essential. Their intermediate enrichment timescales and strong metallicity dependence enable elements such as C, N, Sr, and Ba to probe phases of chemical evolution on shorter timescales. However, doing so requires moving beyond a two-channel enrichment model and placing greater reliance on theoretical yield models. For N and C, \citet{johnson_empirical_2023} and \citet{boyea25} pursue a hybrid theoretical-empirical approach, using AGB models to predict metallicity and time dependence but normalizing the massive star and AGB contributions empirically by fitting Milky Way disk observations. 

% and thus relying on theoretical yields. At present, uncertainties in AGB yield calculations limit the precision of this approach. Improved yields will be required to fully exploit these elements as SFH tracers.

% Their intermediate enrichment timescales and strong metallicity dependence allow elements such as C, N, Sr, and Ba to probe phases of chemical evolution invisible to the CCSN/SNeIa channels. 
% Thus, they offer the most promising path forward for extracting more detailed SFH information from chemical elements.

\subsection{Implications for SPS modeling}

While the limited information content of $\alpha$ and Fe-peak elements is a constraint on SFH recovery, it is in another sense an asset: it means the abundance patterns of quiescent galaxies are highly predictable from just two observables, [Mg/H] and [Mg/Fe]. This predictability has direct implications for stellar population synthesis (SPS) modeling.

Most SPS models assume a solar-scaled abundance pattern --- not because this reflects reality, but because no reliable prescription existed for how individual elements should vary. Recent progress has begun to relax this assumption by allowing variable $\alpha$/Fe ratios \citep[][]{knowles_smiles_2023,park2025_alphaMC}, and these models already reveal substantial effects on colors, mass-to-light ratios, and inferred stellar ages. As we show in this paper, individual $\alpha$ elements do not vary in lockstep: Mg, Si, Ca, and Ti each follow distinct trends with [Mg/Fe] (see Figures~\ref{fig:data} and \ref{fig:deltas}), so a single $\alpha$/Fe parameter only captures a coarse approximation of the true abundance pattern. Our empirical yields provide a natural and physically motivated prescription for how each element varies individually with [Mg/H] and [Mg/Fe], offering a more accurate foundation for the next generation of SPS models.

\section{Conclusion}
\label{sec:conclusion}

We introduce a framework for predicting the multi-element abundance patterns of massive quiescent galaxies using empirically calibrated nucleosynthetic yields derived from Milky Way stars, covering 14 elements. Rather than relying on uncertain theoretical yields or explicit chemical-evolution modeling, we treat each galaxy as a single stellar population whose abundance pattern reflects a mixture of CCSNe and SNe~Ia enrichment. The yields are derived from trends in the APOGEE and GALAH surveys via the two-process framework of \citetalias{weinberg22_apogee}, which decomposes stellar abundances into prompt (CCSN) and delayed (SNe~Ia) components. We apply this approach to quiescent galaxies at $z\sim0$ (SDSS), $z\sim0.7$ (LEGA-C), and $z\sim2$ (\textit{JWST}/SUSPENSE). 

At $z\sim0$, Milky Way calibrated yields reproduce $\alpha$ and Fe-peak abundances to within 0.03~dex, compared to 0.23~dex for commonly used theoretical yield sets. The three theoretical yield sets themselves disagree by up to 0.8~dex, underscoring the magnitude of yield uncertainties. At higher redshifts ($z\sim0.7$ and $z\sim2$), the empirical yields remain accurate to within $\sim$0.06~dex. This consistency across three redshifts and a range of velocity dispersions, combined with prior results in dwarf galaxies and the Milky Way disk, suggests that IMF-averaged nucleosynthetic yields are largely universal across star-forming environments. The framework fails predictably for N, Sr, and Ba, and (at $z\sim2$) C, elements with significant AGB contributions not captured by a two-channel enrichment model. Including an explicit AGB enrichment channel in \textsc{vice} largely resolves the nitrogen discrepancy, confirming that a third enrichment component is required for these elements.

Perhaps the most unexpected result is that incorporating a full metallicity-evolution history offers almost no improvement over anchoring predictions to a single effective metallicity. Most $\alpha$ and Fe-peak yields are nearly metallicity-independent, and Mg and Fe already encode the CCSN-to-SNe~Ia enrichment ratio. As a result, additional $\alpha$ and Fe-peak elements reinforce this measurement rather than providing independent constraints on a galaxy's SFH. For finer SFH detail, AGB-dominated elements such as C, N, Sr, and Ba are promising: their enrichment timescales and metallicity-dependent yields respond to phases of chemical evolution that CCSN/SNeIa channels cannot distinguish.

We also explore the impact of IMF variations. Changing the IMF slope from -2.3 (Kroupa) to -1.9 shifts the theoretical IMF-averaged CCSN abundance ratios [X/Fe]$_{\rm cc}$ by only $\sim$0.05--0.2~dex. The direction of these shifts is consistent with the observed residuals under a top-heavy IMF: enhanced CCSN production suppresses Ba while boosting Sr, matching our findings, and may account for some of the $[\mathrm{Co/Fe}]$ trend with velocity dispersion. Cleanly separating IMF-driven effects from AGB contributions, however, will require a chemical evolution model that handles both simultaneously.

The predictability of multi-element abundance patterns has direct implications for both SPS modeling and cosmological simulations. Current SPS models that allow variable $\alpha$/Fe implicitly assume that all $\alpha$-elements vary in lockstep --- an assumption our results show is incorrect. An element-by-element prescription anchored to [Mg/H] and [Mg/Fe] would provide stronger constraints on stellar ages, mass-to-light ratios, and star-formation histories. More broadly, the success of our empirical yields opens a new avenue for cosmological simulations, which have necessarily relied on theoretical yields. The yield tables provided here (Tables~\ref{tab:yields} and \ref{tab:yields_cnba}) offer an empirically grounded prescription ready for direct incorporation into the chemical enrichment modules of cosmological simulations.

\section*{Acknowledgments}
We thank James Johnson and Charlie Conroy for helpful discussions. AGB acknowledges support from NASA through the NASA Hubble Fellowship grant HST-HF2-51571 awarded by the Space Telescope Science Institute, which is operated by the Association of Universities for Research in Astronomy, Inc., for NASA, under contract NAS5-26555. DHW and NG acknowledge support from NSF grant AST-2307621. MK acknowledges funding from the Dutch Research Council (NWO) through the award of the Vici grant VI.C.222.047 (project 2010007169). 

\software{\texttt{emcee} \citep{foreman-mackey_emcee_2013}, \texttt{alf$\alpha$} \citep{beverage_suspense_2025}, \texttt{fanCE} \citep{gountanis_modeling_2024}, \texttt{VICE} \citep{johnson2020, Johnson21_vice}}

\bibliography{suspense_abundances}{}

@ARTICLE{thoams1998_paper1,
       author = {{Thomas}, D. and {Greggio}, L. and {Bender}, R.},
        title = "{Stellar Yields and Chemical Evolution - I. Abundance Ratios and Delayed Mixing in the Solar Neighbourhood}",
      journal = {\mnras},
         year = 1998,
        month = may,
       volume = {296},
       number = {1},
        pages = {119-149},
          doi = {10.1046/j.1365-8711.1998.01289.x},
archivePrefix = {arXiv},
       eprint = {astro-ph/9710004},
 primaryClass = {astro-ph},
       adsurl = {https://ui.adsabs.harvard.edu/abs/1998MNRAS.296..119T}
}

@ARTICLE{johnson2020,
       author = {{Johnson}, James W. and {Weinberg}, David H.},
        title = "{The impact of starbursts on element abundance ratios}",
      journal = {\mnras},
         year = 2020,
        month = oct,
       volume = {498},
       number = {1},
        pages = {1364-1381},
          doi = {10.1093/mnras/staa2431},
archivePrefix = {arXiv},
       eprint = {1911.02598},
 primaryClass = {astro-ph.GA},
       adsurl = {https://ui.adsabs.harvard.edu/abs/2020MNRAS.498.1364J}
}

@ARTICLE{Seitenzahl2013,
       author = {{Seitenzahl}, Ivo R. and {Ciaraldi-Schoolmann}, Franco and {R{\"o}pke}, Friedrich K. and {Fink}, Michael and {Hillebrandt}, Wolfgang and {Kromer}, Markus and {Pakmor}, R{\"u}diger and {Ruiter}, Ashley J. and {Sim}, Stuart A. and {Taubenberger}, Stefan},
        title = "{Three-dimensional delayed-detonation models with nucleosynthesis for Type Ia supernovae}",
      journal = {\mnras},
         year = 2013,
        month = feb,
       volume = {429},
       number = {2},
        pages = {1156-1172},
          doi = {10.1093/mnras/sts402},
archivePrefix = {arXiv},
       eprint = {1211.3015},
 primaryClass = {astro-ph.SR},
       adsurl = {https://ui.adsabs.harvard.edu/abs/2013MNRAS.429.1156S}
}

@ARTICLE{magg22,
       author = {{Magg}, Ekaterina and {Bergemann}, Maria and {Serenelli}, Aldo and {Bautista}, Manuel and {Plez}, Bertrand and {Heiter}, Ulrike and {Gerber}, Jeffrey M. and {Ludwig}, Hans-G{\"u}nter and {Basu}, Sarbani and {Ferguson}, Jason W. and {Gallego}, Helena Carvajal and {Gamrath}, S{\'e}bastien and {Palmeri}, Patrick and {Quinet}, Pascal},
        title = "{Observational constraints on the origin of the elements. IV. Standard composition of the Sun}",
      journal = {\aap},
         year = 2022,
        month = may,
       volume = {661},
          eid = {A140},
        pages = {A140},
          doi = {10.1051/0004-6361/202142971},
archivePrefix = {arXiv},
       eprint = {2203.02255},
 primaryClass = {astro-ph.SR},
       adsurl = {https://ui.adsabs.harvard.edu/abs/2022A&A...661A.140M}
}

@ARTICLE{lodders25,
       author = {{Lodders}, K. and {Bergemann}, M. and {Palme}, H.},
        title = "{Solar System Elemental Abundances from the Solar Photosphere and CI-Chondrites}",
      journal = {\ssr},
         year = 2025,
        month = mar,
       volume = {221},
       number = {2},
          eid = {23},
        pages = {23},
          doi = {10.1007/s11214-025-01146-w},
archivePrefix = {arXiv},
       eprint = {2502.10575},
 primaryClass = {astro-ph.SR},
       adsurl = {https://ui.adsabs.harvard.edu/abs/2025SSRv..221...23L}
}

@ARTICLE{chieffi13,
       author = {{Chieffi}, Alessandro and {Limongi}, Marco},
        title = "{Pre-supernova Evolution of Rotating Solar Metallicity Stars in the Mass Range 13-120 M $_{☉}$ and their Explosive Yields}",
      journal = {\apj},
         year = 2013,
        month = feb,
       volume = {764},
       number = {1},
          eid = {21},
        pages = {21},
          doi = {10.1088/0004-637X/764/1/21},
       adsurl = {https://ui.adsabs.harvard.edu/abs/2013ApJ...764...21C}
}

@ARTICLE{boyea25,
       author = {{Boyea}, Daniel A. and {Johnson}, James W. and {Weinberg}, David H.},
        title = "{The galactic chemical evolution of carbon: Implications for stellar nucleosynthesis}",
      journal = {arXiv e-prints},
         year = 2025,
        month = nov,
          eid = {arXiv:2511.20752},
        pages = {arXiv:2511.20752},
          doi = {10.48550/arXiv.2511.20752},
archivePrefix = {arXiv},
       eprint = {2511.20752},
 primaryClass = {astro-ph.GA},
       adsurl = {https://ui.adsabs.harvard.edu/abs/2025arXiv251120752B}
}

@ARTICLE{guo24,
       author = {{Guo}, Ziyi and {Zhang}, Zhi-Yu and {Yan}, Zhiqiang and {Gjergo}, Eda and {Man}, Allison W.~S. and {Ivison}, R.~J. and {Fu}, Xiaoting and {Shi}, Yong},
        title = "{First Detection of CO Isotopologues in a High-redshift Main-sequence Galaxy: Evidence of a Top-heavy Stellar Initial Mass Function}",
      journal = {\apj},
         year = 2024,
        month = aug,
       volume = {970},
       number = {2},
          eid = {136},
        pages = {136},
          doi = {10.3847/1538-4357/ad4da2},
archivePrefix = {arXiv},
       eprint = {2405.05317},
 primaryClass = {astro-ph.GA},
       adsurl = {https://ui.adsabs.harvard.edu/abs/2024ApJ...970..136G}
}

@ARTICLE{zhang18,
       author = {{Zhang}, Zhi-Yu and {Romano}, D. and {Ivison}, R.~J. and {Papadopoulos}, Padelis P. and {Matteucci}, F.},
        title = "{Stellar populations dominated by massive stars in dusty starburst galaxies across cosmic time}",
      journal = {\nat},
         year = 2018,
        month = jun,
       volume = {558},
       number = {7709},
        pages = {260-263},
          doi = {10.1038/s41586-018-0196-x},
archivePrefix = {arXiv},
       eprint = {1806.01280},
 primaryClass = {astro-ph.GA},
       adsurl = {https://ui.adsabs.harvard.edu/abs/2018Natur.558..260Z}
}

@ARTICLE{griffith24,
       author = {{Griffith}, Emily J. and {Hogg}, David W. and {Dalcanton}, Julianne J. and {Hasselquist}, Sten and {Ratcliffe}, Bridget and {Ness}, Melissa and {Weinberg}, David H.},
        title = "{KPM: A Flexible and Data-driven K-process Model for Nucleosynthesis}",
      journal = {\aj},
         year = 2024,
        month = mar,
       volume = {167},
       number = {3},
          eid = {98},
        pages = {98},
          doi = {10.3847/1538-3881/ad19c7},
archivePrefix = {arXiv},
       eprint = {2307.05691},
 primaryClass = {astro-ph.GA},
       adsurl = {https://ui.adsabs.harvard.edu/abs/2024AJ....167...98G}
}

@ARTICLE{sit25,
       author = {{Sit}, Tawny and {Weinberg}, David H. and {Griffith}, Emily J.},
        title = "{On the Origin of Abundance Variations in the Milky Way's High-{\ensuremath{\alpha}} Plateau}",
      journal = {\apj},
         year = 2025,
        month = nov,
       volume = {994},
       number = {1},
          eid = {53},
        pages = {53},
          doi = {10.3847/1538-4357/ae0b5d},
archivePrefix = {arXiv},
       eprint = {2503.07738},
 primaryClass = {astro-ph.GA},
       adsurl = {https://ui.adsabs.harvard.edu/abs/2025ApJ...994...53S}
}

@ARTICLE{hasselquist24,
       author = {{Hasselquist}, Sten and {Hayes}, Christian R. and {Griffith}, Emily J. and {Weinberg}, David and {Sit}, Tawny and {Beaton}, Rachael L. and {Horta}, Danny},
        title = "{Two-process Model and Residual Abundance Analysis of the Milky Way Massive Satellites}",
      journal = {\apj},
         year = 2024,
        month = oct,
       volume = {974},
       number = {2},
          eid = {227},
        pages = {227},
          doi = {10.3847/1538-4357/ad70ad},
archivePrefix = {arXiv},
       eprint = {2408.10393},
 primaryClass = {astro-ph.GA},
       adsurl = {https://ui.adsabs.harvard.edu/abs/2024ApJ...974..227H}
}

@ARTICLE{sanders2025,
       author = {{Sanders}, Jason L.},
        title = "{Chemical separation of stellar populations: analytic solutions for chemical evolution models with metallicity-dependent yields}",
      journal = {\mnras},
         year = 2025,
        month = dec,
       volume = {544},
       number = {4},
        pages = {4590-4610},
          doi = {10.1093/mnras/staf1900},
archivePrefix = {arXiv},
       eprint = {2510.25876},
 primaryClass = {astro-ph.GA},
       adsurl = {https://ui.adsabs.harvard.edu/abs/2025MNRAS.544.4590S}
}

@ARTICLE{Cohen2004,
       author = {{Cohen}, Judith G. and {Christlieb}, Norbert and {McWilliam}, Andrew and {Shectman}, Steve and {Thompson}, Ian and {Wasserburg}, G.~J. and {Ivans}, Inese and {Dehn}, Matthias and {Karlsson}, Torgny and {Melendez}, J.},
        title = "{Abundances In Very Metal-Poor Dwarf Stars}",
      journal = {\apj},
         year = 2004,
        month = sep,
       volume = {612},
       number = {2},
        pages = {1107-1135},
          doi = {10.1086/422576},
archivePrefix = {arXiv},
       eprint = {astro-ph/0405286},
 primaryClass = {astro-ph},
       adsurl = {https://ui.adsabs.harvard.edu/abs/2004ApJ...612.1107C}
}

@software{alfsoft2023,
       author = {{Conroy}, Charlie and {van Dokkum}, Pieter and {Villaume}, Alexa and {Lind}, Karin},
        title = "{ALF: Absorption line fitter}",
 howpublished = {Astrophysics Source Code Library, record ascl:2307.004},
         year = 2023,
        month = jul,
          eid = {ascl:2307.004},
archivePrefix = {ascl},
       eprint = {2307.004},
       adsurl = {https://ui.adsabs.harvard.edu/abs/2023ascl.soft07004C}
}

@ARTICLE{majewski2017,
       author = {{Majewski}, Steven R. and {Schiavon}, Ricardo P. and {Frinchaboy}, Peter M. and {Allende Prieto}, Carlos and {Barkhouser}, Robert and {Bizyaev}, Dmitry and {Blank}, Basil and {Brunner}, Sophia and {Burton}, Adam and {Carrera}, Ricardo and {Chojnowski}, S. Drew and {Cunha}, K{\'a}tia and {Epstein}, Courtney and {Fitzgerald}, Greg and {Garc{\'\i}a P{\'e}rez}, Ana E. and {Hearty}, Fred R. and {Henderson}, Chuck and {Holtzman}, Jon A. and {Johnson}, Jennifer A. and {Lam}, Charles R. and {Lawler}, James E. and {Maseman}, Paul and {M{\'e}sz{\'a}ros}, Szabolcs and {Nelson}, Matthew and {Nguyen}, Duy Coung and {Nidever}, David L. and {Pinsonneault}, Marc and {Shetrone}, Matthew and {Smee}, Stephen and {Smith}, Verne V. and {Stolberg}, Todd and {Skrutskie}, Michael F. and {Walker}, Eric and {Wilson}, John C. and {Zasowski}, Gail and {Anders}, Friedrich and {Basu}, Sarbani and {Beland}, Stephane and {Blanton}, Michael R. and {Bovy}, Jo and {Brownstein}, Joel R. and {Carlberg}, Joleen and {Chaplin}, William and {Chiappini}, Cristina and {Eisenstein}, Daniel J. and {Elsworth}, Yvonne and {Feuillet}, Diane and {Fleming}, Scott W. and {Galbraith-Frew}, Jessica and {Garc{\'\i}a}, Rafael A. and {Garc{\'\i}a-Hern{\'a}ndez}, D. An{\'\i}bal and {Gillespie}, Bruce A. and {Girardi}, L{\'e}o and {Gunn}, James E. and {Hasselquist}, Sten and {Hayden}, Michael R. and {Hekker}, Saskia and {Ivans}, Inese and {Kinemuchi}, Karen and {Klaene}, Mark and {Mahadevan}, Suvrath and {Mathur}, Savita and {Mosser}, Beno{\^\i}t and {Muna}, Demitri and {Munn}, Jeffrey A. and {Nichol}, Robert C. and {O'Connell}, Robert W. and {Parejko}, John K. and {Robin}, A.~C. and {Rocha-Pinto}, Helio and {Schultheis}, Matthias and {Serenelli}, Aldo M. and {Shane}, Neville and {Silva Aguirre}, Victor and {Sobeck}, Jennifer S. and {Thompson}, Benjamin and {Troup}, Nicholas W. and {Weinberg}, David H. and {Zamora}, Olga},
        title = "{The Apache Point Observatory Galactic Evolution Experiment (APOGEE)}",
      journal = {\aj},
         year = 2017,
        month = sep,
       volume = {154},
       number = {3},
          eid = {94},
        pages = {94},
          doi = {10.3847/1538-3881/aa784d},
archivePrefix = {arXiv},
       eprint = {1509.05420},
 primaryClass = {astro-ph.IM},
       adsurl = {https://ui.adsabs.harvard.edu/abs/2017AJ....154...94M}
}

@ARTICLE{Maeder2012,
       author = {{Maeder}, Andr{\'e} and {Meynet}, Georges},
        title = "{Rotating massive stars: From first stars to gamma ray bursts}",
      journal = {Reviews of Modern Physics},
         year = 2012,
        month = jan,
       volume = {84},
       number = {1},
        pages = {25-63},
          doi = {10.1103/RevModPhys.84.25},
       adsurl = {https://ui.adsabs.harvard.edu/abs/2012RvMP...84...25M}
}

@ARTICLE{Maeder1992,
       author = {{Maeder}, Andre},
        title = "{Stellar yields as a function of initial metallicity and mass limit for black hole formation}",
      journal = {\aap},
         year = 1992,
        month = oct,
       volume = {264},
       number = {1},
        pages = {105-120},
       adsurl = {https://ui.adsabs.harvard.edu/abs/1992A&A...264..105M}
}

@ARTICLE{demink2013,
       author = {{de Mink}, S.~E. and {Langer}, N. and {Izzard}, R.~G. and {Sana}, H. and {de Koter}, A.},
        title = "{The Rotation Rates of Massive Stars: The Role of Binary Interaction through Tides, Mass Transfer, and Mergers}",
      journal = {\apj},
         year = 2013,
        month = feb,
       volume = {764},
       number = {2},
          eid = {166},
        pages = {166},
          doi = {10.1088/0004-637X/764/2/166},
archivePrefix = {arXiv},
       eprint = {1211.3742},
 primaryClass = {astro-ph.SR},
       adsurl = {https://ui.adsabs.harvard.edu/abs/2013ApJ...764..166D}
}

@ARTICLE{marchant2024,
       author = {{Marchant}, Pablo and {Bodensteiner}, Julia},
        title = "{The Evolution of Massive Binary Stars}",
      journal = {\araa},
         year = 2024,
        month = sep,
       volume = {62},
       number = {1},
        pages = {21-61},
          doi = {10.1146/annurev-astro-052722-105936},
archivePrefix = {arXiv},
       eprint = {2311.01865},
 primaryClass = {astro-ph.SR},
       adsurl = {https://ui.adsabs.harvard.edu/abs/2024ARA&A..62...21M}
}

@ARTICLE{vink2022,
       author = {{Vink}, Jorick S.},
        title = "{Theory and Diagnostics of Hot Star Mass Loss}",
      journal = {\araa},
         year = 2022,
        month = aug,
       volume = {60},
        pages = {203-246},
          doi = {10.1146/annurev-astro-052920-094949},
archivePrefix = {arXiv},
       eprint = {2109.08164},
 primaryClass = {astro-ph.SR},
       adsurl = {https://ui.adsabs.harvard.edu/abs/2022ARA&A..60..203V}
}

@ARTICLE{huscher2025,
       author = {{Huscher}, Ezra and {Finlator}, Kristian and {Jackiewicz}, Jason},
        title = "{Asymptotic Giant Branch Mass-loss Rates and Metal Yields from Scaled Mixing-length and Mass-loss Parameters}",
      journal = {\apj},
         year = 2025,
        month = nov,
       volume = {993},
       number = {1},
          eid = {16},
        pages = {16},
          doi = {10.3847/1538-4357/ae0199},
       adsurl = {https://ui.adsabs.harvard.edu/abs/2025ApJ...993...16H}
}

@ARTICLE{josiek2024,
       author = {{Josiek}, J. and {Ekstr{\"o}m}, S. and {Sander}, A.~A.~C.},
        title = "{Impact of main sequence mass loss on the appearance, structure, and evolution of Wolf-Rayet stars}",
      journal = {\aap},
         year = 2024,
        month = aug,
       volume = {688},
          eid = {A71},
        pages = {A71},
          doi = {10.1051/0004-6361/202449281},
archivePrefix = {arXiv},
       eprint = {2404.14488},
 primaryClass = {astro-ph.SR},
       adsurl = {https://ui.adsabs.harvard.edu/abs/2024A&A...688A..71J}
}

@ARTICLE{leung2026arXiv,
       author = {{Leung}, Ho-Hin and {Carnall}, Adam C. and {Taylor}, Elizabeth and {Stevenson}, Struan D. and {Beverage}, Aliza G. and {Cullen}, Fergus and {Dunlop}, James S. and {McLeod}, Derek J. and {McLure}, Ross J. and {Begley}, Ryan and {Almaini}, Omar and {Antonogiannaki}, Stella and {Arellano-C{\'o}rdova}, Karla Z. and {Barrufet}, Laia and {Bondestam}, Cecilia and {Donnan}, Callum T. and {Holst}, Isaac J.~B. and {Liu}, Feng-Yuan F. and {Rowlands}, Kate and {Sanders}, Ryan L. and {Scholte}, Dirk and {Skarbinski}, Maya and {Stanton}, Thomas M. and {Wild}, Vivienne},
        title = "{The JWST EXCELS survey: The ages and abundances of $3<z<5$ massive quiescent galaxies show that downsizing was already in place by $z\simeq4$}",
      journal = {arXiv e-prints},
         year = 2026,
        month = feb,
          eid = {arXiv:2602.05934},
        pages = {arXiv:2602.05934},
          doi = {10.48550/arXiv.2602.05934},
archivePrefix = {arXiv},
       eprint = {2602.05934},
 primaryClass = {astro-ph.GA},
       adsurl = {https://ui.adsabs.harvard.edu/abs/2026arXiv260205934L}
}

@ARTICLE{hamadouche2026arXiv,
       author = {{Hamadouche}, Massissilia L. and {Whitaker}, Katherine E. and {Valentino}, Francesco and {Antwi-Danso}, Jacqueline and {Ito}, Kei and {Beverage}, Aliza and {Zhu}, Pengpei and {Brammer}, Gabriel and {Kokorev}, Vasily and {de Lucia}, Gabriella and {Baker}, William M. and {Farcy}, Marion and {Gallazzi}, Anna and {Gillman}, Steven and {Gottumukkala}, Rashmi and {Hirschmann}, Michaela and {Jespersen}, Christian Kragh and {Kakimoto}, Takumi and {Lee}, Minju M. and {Onodera}, Masato and {Shimakawa}, Rhythm and {Tanaka}, Masayuki and {Weaver}, John R. and {Wu}, Po-Feng},
        title = "{DeepDive: Tracing the early quenching pathways of massive quiescent galaxies at $z>3$ from their star-formation histories and chemical abundances}",
      journal = {arXiv e-prints},
         year = 2026,
        month = feb,
          eid = {arXiv:2602.02485},
        pages = {arXiv:2602.02485},
          doi = {10.48550/arXiv.2602.02485},
archivePrefix = {arXiv},
       eprint = {2602.02485},
 primaryClass = {astro-ph.GA},
       adsurl = {https://ui.adsabs.harvard.edu/abs/2026arXiv260202485H}
}

@ARTICLE{tacchella2017,
       author = {{Tacchella}, Sandro and {Carollo}, C. Marcella and {Faber}, S.~M. and {Cibinel}, Anna and {Dekel}, Avishai and {Koo}, David C. and {Renzini}, Alvio and {Woo}, Joanna},
        title = "{On the Evolution of the Central Density of Quiescent Galaxies}",
      journal = {\apjl},
         year = 2017,
        month = jul,
       volume = {844},
       number = {1},
          eid = {L1},
        pages = {L1},
          doi = {10.3847/2041-8213/aa7cfb},
archivePrefix = {arXiv},
       eprint = {1707.00695},
 primaryClass = {astro-ph.GA},
       adsurl = {https://ui.adsabs.harvard.edu/abs/2017ApJ...844L...1T}
}

@ARTICLE{park2025_alphaMC,
       author = {{Park}, Minjung and {Conroy}, Charlie and {Johnson}, Benjamin D. and {Leja}, Joel and {Dotter}, Aaron and {Cargile}, Phillip A.},
        title = "{{\ensuremath{\alpha}}-MC: Self-consistent {\ensuremath{\alpha}}-enhanced Stellar Population Models Covering a Wide Range of Age, Metallicity, and Wavelength}",
      journal = {\apj},
         year = 2025,
        month = dec,
       volume = {994},
       number = {2},
          eid = {165},
        pages = {165},
          doi = {10.3847/1538-4357/ae0cba},
archivePrefix = {arXiv},
       eprint = {2410.21375},
 primaryClass = {astro-ph.GA},
       adsurl = {https://ui.adsabs.harvard.edu/abs/2025ApJ...994..165P}
}

@ARTICLE{bunker2023,
       author = {{Bunker}, Andrew J. and {Saxena}, Aayush and {Cameron}, Alex J. and {Willott}, Chris J. and {Curtis-Lake}, Emma and {Jakobsen}, Peter and {Carniani}, Stefano and {Smit}, Renske and {Maiolino}, Roberto and {Witstok}, Joris and {Curti}, Mirko and {D'Eugenio}, Francesco and {Jones}, Gareth C. and {Ferruit}, Pierre and {Arribas}, Santiago and {Charlot}, Stephane and {Chevallard}, Jacopo and {Giardino}, Giovanna and {de Graaff}, Anna and {Looser}, Tobias J. and {L{\"u}tzgendorf}, Nora and {Maseda}, Michael V. and {Rawle}, Tim and {Rix}, Hans-Walter and {Del Pino}, Bruno Rodr{\'\i}guez and {Alberts}, Stacey and {Egami}, Eiichi and {Eisenstein}, Daniel J. and {Endsley}, Ryan and {Hainline}, Kevin and {Hausen}, Ryan and {Johnson}, Benjamin D. and {Rieke}, George and {Rieke}, Marcia and {Robertson}, Brant E. and {Shivaei}, Irene and {Stark}, Daniel P. and {Sun}, Fengwu and {Tacchella}, Sandro and {Tang}, Mengtao and {Williams}, Christina C. and {Willmer}, Christopher N.~A. and {Baker}, William M. and {Baum}, Stefi and {Bhatawdekar}, Rachana and {Bowler}, Rebecca and {Boyett}, Kristan and {Chen}, Zuyi and {Circosta}, Chiara and {Helton}, Jakob M. and {Ji}, Zhiyuan and {Kumari}, Nimisha and {Lyu}, Jianwei and {Nelson}, Erica and {Parlanti}, Eleonora and {Perna}, Michele and {Sandles}, Lester and {Scholtz}, Jan and {Suess}, Katherine A. and {Topping}, Michael W. and {{\"U}bler}, Hannah and {Wallace}, Imaan E.~B. and {Whitler}, Lily},
        title = "{JADES NIRSpec Spectroscopy of GN-z11: Lyman-{\ensuremath{\alpha}} emission and possible enhanced nitrogen abundance in a z = 10.60 luminous galaxy}",
      journal = {\aap},
         year = 2023,
        month = sep,
       volume = {677},
          eid = {A88},
        pages = {A88},
          doi = {10.1051/0004-6361/202346159},
archivePrefix = {arXiv},
       eprint = {2302.07256},
 primaryClass = {astro-ph.GA},
       adsurl = {https://ui.adsabs.harvard.edu/abs/2023A&A...677A..88B}
}

@ARTICLE{griffith2021,
       author = {{Griffith}, Emily J. and {Sukhbold}, Tuguldur and {Weinberg}, David H. and {Johnson}, Jennifer A. and {Johnson}, James W. and {Vincenzo}, Fiorenzo},
        title = "{The Impact of Black Hole Formation on Population-averaged Supernova Yields}",
      journal = {\apj},
         year = 2021,
        month = nov,
       volume = {921},
       number = {1},
          eid = {73},
        pages = {73},
          doi = {10.3847/1538-4357/ac1bac},
archivePrefix = {arXiv},
       eprint = {2103.09837},
 primaryClass = {astro-ph.SR},
       adsurl = {https://ui.adsabs.harvard.edu/abs/2021ApJ...921...73G}
}

@ARTICLE{ertl2016,
       author = {{Ertl}, T. and {Janka}, H.-Th. and {Woosley}, S.~E. and {Sukhbold}, T. and {Ugliano}, M.},
        title = "{A Two-parameter Criterion for Classifying the Explodability of Massive Stars by the Neutrino-driven Mechanism}",
      journal = {\apj},
         year = 2016,
        month = feb,
       volume = {818},
       number = {2},
          eid = {124},
        pages = {124},
          doi = {10.3847/0004-637X/818/2/124},
archivePrefix = {arXiv},
       eprint = {1503.07522},
 primaryClass = {astro-ph.SR},
       adsurl = {https://ui.adsabs.harvard.edu/abs/2016ApJ...818..124E}
}

@ARTICLE{sukhbold2014,
       author = {{Sukhbold}, Tuguldur and {Woosley}, S.~E.},
        title = "{The Compactness of Presupernova Stellar Cores}",
      journal = {\apj},
         year = 2014,
        month = mar,
       volume = {783},
       number = {1},
          eid = {10},
        pages = {10},
          doi = {10.1088/0004-637X/783/1/10},
archivePrefix = {arXiv},
       eprint = {1311.6546},
 primaryClass = {astro-ph.SR},
       adsurl = {https://ui.adsabs.harvard.edu/abs/2014ApJ...783...10S}
}

@ARTICLE{trinca2024,
       author = {{Trinca}, Alessandro and {Schneider}, Raffaella and {Valiante}, Rosa and {Graziani}, Luca and {Ferrotti}, Arianna and {Omukai}, Kazuyuki and {Chon}, Sunmyon},
        title = "{Exploring the nature of UV-bright z {\ensuremath{\gtrsim}} 10 galaxies detected by JWST: star formation, black hole accretion, or a non-universal IMF?}",
      journal = {\mnras},
         year = 2024,
        month = apr,
       volume = {529},
       number = {4},
        pages = {3563-3581},
          doi = {10.1093/mnras/stae651},
archivePrefix = {arXiv},
       eprint = {2305.04944},
 primaryClass = {astro-ph.GA},
       adsurl = {https://ui.adsabs.harvard.edu/abs/2024MNRAS.529.3563T}
}

@ARTICLE{Ziparo2023,
       author = {{Ziparo}, Francesco and {Ferrara}, Andrea and {Sommovigo}, Laura and {Kohandel}, Mahsa},
        title = "{Blue monsters. Why are JWST super-early, massive galaxies so blue?}",
      journal = {\mnras},
         year = 2023,
        month = apr,
       volume = {520},
       number = {2},
        pages = {2445-2450},
          doi = {10.1093/mnras/stad125},
archivePrefix = {arXiv},
       eprint = {2209.06840},
 primaryClass = {astro-ph.GA},
       adsurl = {https://ui.adsabs.harvard.edu/abs/2023MNRAS.520.2445Z}
}

@ARTICLE{Ferrara2025,
       author = {{Ferrara}, A. and {Pallottini}, A. and {Sommovigo}, L.},
        title = "{Blue monsters at z > 10: Where all their dust has gone}",
      journal = {\aap},
         year = 2025,
        month = feb,
       volume = {694},
          eid = {A286},
        pages = {A286},
          doi = {10.1051/0004-6361/202452707},
archivePrefix = {arXiv},
       eprint = {2410.19042},
 primaryClass = {astro-ph.GA},
       adsurl = {https://ui.adsabs.harvard.edu/abs/2025A&A...694A.286F}
}

@ARTICLE{Hutter2025,
       author = {{Hutter}, Anne and {Cueto}, Elie R. and {Dayal}, Pratika and {Gottl{\"o}ber}, Stefan and {Trebitsch}, Maxime and {Yepes}, Gustavo},
        title = "{ASTRAEUS: X. Indications of a top-heavy initial mass function in highly star-forming galaxies from JWST observations at z > 10}",
      journal = {\aap},
         year = 2025,
        month = feb,
       volume = {694},
          eid = {A254},
        pages = {A254},
          doi = {10.1051/0004-6361/202452460},
archivePrefix = {arXiv},
       eprint = {2410.00730},
 primaryClass = {astro-ph.GA},
       adsurl = {https://ui.adsabs.harvard.edu/abs/2025A&A...694A.254H}
}

@ARTICLE{Mcleod2024,
       author = {{McLeod}, D.~J. and {Donnan}, C.~T. and {McLure}, R.~J. and {Dunlop}, J.~S. and {Magee}, D. and {Begley}, R. and {Carnall}, A.~C. and {Cullen}, F. and {Ellis}, R.~S. and {Hamadouche}, M.~L. and {Stanton}, T.~M.},
        title = "{The galaxy UV luminosity function at z ≃ 11 from a suite of public JWST ERS, ERO, and Cycle-1 programs}",
      journal = {\mnras},
         year = 2024,
        month = jan,
       volume = {527},
       number = {3},
        pages = {5004-5022},
          doi = {10.1093/mnras/stad3471},
archivePrefix = {arXiv},
       eprint = {2304.14469},
 primaryClass = {astro-ph.GA},
       adsurl = {https://ui.adsabs.harvard.edu/abs/2024MNRAS.527.5004M}
}

@ARTICLE{Harikane2023,
       author = {{Harikane}, Yuichi and {Ouchi}, Masami and {Oguri}, Masamune and {Ono}, Yoshiaki and {Nakajima}, Kimihiko and {Isobe}, Yuki and {Umeda}, Hiroya and {Mawatari}, Ken and {Zhang}, Yechi},
        title = "{A Comprehensive Study of Galaxies at z   9-16 Found in the Early JWST Data: Ultraviolet Luminosity Functions and Cosmic Star Formation History at the Pre-reionization Epoch}",
      journal = {\apjs},
         year = 2023,
        month = mar,
       volume = {265},
       number = {1},
          eid = {5},
        pages = {5},
          doi = {10.3847/1538-4365/acaaa9},
archivePrefix = {arXiv},
       eprint = {2208.01612},
 primaryClass = {astro-ph.GA},
       adsurl = {https://ui.adsabs.harvard.edu/abs/2023ApJS..265....5H}
}

@ARTICLE{finkelstein2024,
       author = {{Finkelstein}, Steven L. and {Leung}, Gene C.~K. and {Bagley}, Micaela B. and {Dickinson}, Mark and {Ferguson}, Henry C. and {Papovich}, Casey and {Akins}, Hollis B. and {Arrabal Haro}, Pablo and {Dav{\'e}}, Romeel and {Dekel}, Avishai and {Kartaltepe}, Jeyhan S. and {Kocevski}, Dale D. and {Koekemoer}, Anton M. and {Pirzkal}, Nor and {Somerville}, Rachel S. and {Yung}, L.~Y. Aaron and {Amor{\'\i}n}, Ricardo O. and {Backhaus}, Bren E. and {Behroozi}, Peter and {Bisigello}, Laura and {Bromm}, Volker and {Casey}, Caitlin M. and {Ch{\'a}vez Ortiz}, {\'O}scar A. and {Cheng}, Yingjie and {Chworowsky}, Katherine and {Cleri}, Nikko J. and {Cooper}, M.~C. and {Davis}, Kelcey and {de la Vega}, Alexander and {Elbaz}, David and {Franco}, Maximilien and {Fontana}, Adriano and {Fujimoto}, Seiji and {Giavalisco}, Mauro and {Grogin}, Norman A. and {Holwerda}, Benne W. and {Huertas-Company}, Marc and {Hirschmann}, Michaela and {Iyer}, Kartheik G. and {Jogee}, Shardha and {Jung}, Intae and {Larson}, Rebecca L. and {Lucas}, Ray A. and {Mobasher}, Bahram and {Morales}, Alexa M. and {Morley}, Caroline V. and {Mukherjee}, Sagnick and {P{\'e}rez-Gonz{\'a}lez}, Pablo G. and {Ravindranath}, Swara and {Rodighiero}, Giulia and {Rowland}, Melanie J. and {Tacchella}, Sandro and {Taylor}, Anthony J. and {Trump}, Jonathan R. and {Wilkins}, Stephen M.},
        title = "{The Complete CEERS Early Universe Galaxy Sample: A Surprisingly Slow Evolution of the Space Density of Bright Galaxies at z {\ensuremath{\sim}} 8.5{\textendash}14.5}",
      journal = {\apjl},
         year = 2024,
        month = jul,
       volume = {969},
       number = {1},
          eid = {L2},
        pages = {L2},
          doi = {10.3847/2041-8213/ad4495},
archivePrefix = {arXiv},
       eprint = {2311.04279},
 primaryClass = {astro-ph.GA},
       adsurl = {https://ui.adsabs.harvard.edu/abs/2024ApJ...969L...2F}
}

@ARTICLE{conroy22_h3,
       author = {{Conroy}, Charlie and {Weinberg}, David H. and {Naidu}, Rohan P. and {Buck}, Tobias and {Johnson}, James W. and {Cargile}, Phillip and {Bonaca}, Ana and {Caldwell}, Nelson and {Chandra}, Vedant and {Han}, Jiwon Jesse and {Johnson}, Benjamin D. and {Speagle}, Joshua S. and {Ting}, Yuan-Sen and {Woody}, Turner and {Zaritsky}, Dennis},
        title = "{Birth of the Galactic Disk Revealed by the H3 Survey}",
      journal = {arXiv e-prints},
         year = 2022,
        month = apr,
          eid = {arXiv:2204.02989},
        pages = {arXiv:2204.02989},
          doi = {10.48550/arXiv.2204.02989},
archivePrefix = {arXiv},
       eprint = {2204.02989},
 primaryClass = {astro-ph.GA},
       adsurl = {https://ui.adsabs.harvard.edu/abs/2022arXiv220402989C}
}

@ARTICLE{Rodriguez23_SNe,
       author = {{Rodr{\'\i}guez}, {\'O}smar and {Maoz}, Dan and {Nakar}, Ehud},
        title = "{The Iron Yield of Core-collapse Supernovae}",
      journal = {\apj},
         year = 2023,
        month = sep,
       volume = {955},
       number = {1},
          eid = {71},
        pages = {71},
          doi = {10.3847/1538-4357/ace2bd},
archivePrefix = {arXiv},
       eprint = {2209.05552},
 primaryClass = {astro-ph.HE},
       adsurl = {https://ui.adsabs.harvard.edu/abs/2023ApJ...955...71R}
}

@ARTICLE{Johnson21_vice,
       author = {{Johnson}, James W. and {Weinberg}, David H. and {Vincenzo}, Fiorenzo and {Bird}, Jonathan C. and {Loebman}, Sarah R. and {Brooks}, Alyson M. and {Quinn}, Thomas R. and {Christensen}, Charlotte R. and {Griffith}, Emily J.},
        title = "{Stellar migration and chemical enrichment in the milky way disc: a hybrid model}",
      journal = {\mnras},
         year = 2021,
        month = dec,
       volume = {508},
       number = {3},
        pages = {4484-4511},
          doi = {10.1093/mnras/stab2718},
archivePrefix = {arXiv},
       eprint = {2103.09838},
 primaryClass = {astro-ph.GA},
       adsurl = {https://ui.adsabs.harvard.edu/abs/2021MNRAS.508.4484J}
}

@ARTICLE{weinberg24_yields,
       author = {{Weinberg}, David H. and {Griffith}, Emily J. and {Johnson}, James W. and {Thompson}, Todd A.},
        title = "{The Scale of Stellar Yields: Implications of the Measured Mean Iron Yield of Core Collapse Supernovae}",
      journal = {\apj},
         year = 2024,
        month = oct,
       volume = {973},
       number = {2},
          eid = {122},
        pages = {122},
          doi = {10.3847/1538-4357/ad6313},
archivePrefix = {arXiv},
       eprint = {2309.05719},
 primaryClass = {astro-ph.GA},
       adsurl = {https://ui.adsabs.harvard.edu/abs/2024ApJ...973..122W}
}

@ARTICLE{griffith19_galah,
       author = {{Griffith}, Emily and {Johnson}, Jennifer A. and {Weinberg}, David H.},
        title = "{Abundance Ratios in GALAH DR2 and Their Implications for Nucleosynthesis}",
      journal = {\apj},
         year = 2019,
        month = dec,
       volume = {886},
       number = {2},
          eid = {84},
        pages = {84},
          doi = {10.3847/1538-4357/ab4b5d},
archivePrefix = {arXiv},
       eprint = {1908.06113},
 primaryClass = {astro-ph.SR},
       adsurl = {https://ui.adsabs.harvard.edu/abs/2019ApJ...886...84G}
}

@ARTICLE{weinberg22_apogee,
       author = {{Weinberg}, David H. and {Holtzman}, Jon A. and {Johnson}, Jennifer A. and {Hayes}, Christian and {Hasselquist}, Sten and {Shetrone}, Matthew and {Ting}, Yuan-Sen and {Beaton}, Rachael L. and {Beers}, Timothy C. and {Bird}, Jonathan C. and {Bizyaev}, Dmitry and {Blanton}, Michael R. and {Cunha}, Katia and {Fern{\'a}ndez-Trincado}, Jos{\'e} G. and {Frinchaboy}, Peter M. and {Garc{\'\i}a-Hern{\'a}ndez}, D.~A. and {Griffith}, Emily and {Johnson}, James W. and {J{\"o}nsson}, Henrik and {Lane}, Richard R. and {Leung}, Henry W. and {Mackereth}, J. Ted and {Majewski}, Steven R. and {M{\'e}sz{\'a}ros}, Szabolcs and {Nitschelm}, Christian and {Pan}, Kaike and {Schiavon}, Ricardo P. and {Schneider}, Donald P. and {Schultheis}, Mathias and {Smith}, Verne and {Sobeck}, Jennifer S. and {Stassun}, Keivan G. and {Stringfellow}, Guy S. and {Vincenzo}, Fiorenzo and {Wilson}, John C. and {Zasowski}, Gail},
        title = "{Chemical Cartography with APOGEE: Mapping Disk Populations with a 2-process Model and Residual Abundances}",
      journal = {\apjs},
         year = 2022,
        month = jun,
       volume = {260},
       number = {2},
          eid = {32},
        pages = {32},
          doi = {10.3847/1538-4365/ac6028},
archivePrefix = {arXiv},
       eprint = {2108.08860},
 primaryClass = {astro-ph.GA},
       adsurl = {https://ui.adsabs.harvard.edu/abs/2022ApJS..260...32W}
}

@ARTICLE{WAF2017,
       author = {{Weinberg}, David H. and {Andrews}, Brett H. and {Freudenburg}, Jenna},
        title = "{Equilibrium and Sudden Events in Chemical Evolution}",
      journal = {\apj},
         year = 2017,
        month = mar,
       volume = {837},
       number = {2},
          eid = {183},
        pages = {183},
          doi = {10.3847/1538-4357/837/2/183},
archivePrefix = {arXiv},
       eprint = {1604.07435},
 primaryClass = {astro-ph.GA},
       adsurl = {https://ui.adsabs.harvard.edu/abs/2017ApJ...837..183W}
}

@ARTICLE{beverage_suspense_2025,
       author = {{Beverage}, Aliza G. and {Slob}, Martje and {Kriek}, Mariska and {Conroy}, Charlie and {Barro}, Guillermo and {Bezanson}, Rachel and {Brammer}, Gabriel and {Cheng}, Chloe M. and {de Graaff}, Anna and {F{\"o}rster Schreiber}, Natascha M. and {Franx}, Marijn and {Lorenz}, Brian and {Mancera Pi{\~n}a}, Pavel E. and {Marchesini}, Danilo and {Muzzin}, Adam and {Newman}, Andrew B. and {Price}, Sedona H. and {Shapley}, Alice E. and {Stefanon}, Mauro and {Suess}, Katherine A. and {van Dokkum}, Pieter and {Weinberg}, David and {Weisz}, Daniel R.},
        title = "{Carbon and Iron Deficiencies in Quiescent Galaxies at z = 1{\textendash}3 from JWST-SUSPENSE: Implications for the Formation Histories of Massive Galaxies}",
      journal = {\apj},
         year = 2025,
        month = feb,
       volume = {979},
       number = {2},
          eid = {249},
        pages = {249},
          doi = {10.3847/1538-4357/ad96b6},
archivePrefix = {arXiv},
       eprint = {2407.02556},
 primaryClass = {astro-ph.GA},
       adsurl = {https://ui.adsabs.harvard.edu/abs/2025ApJ...979..249B}
}

@article{denbrok_recovery_2024,
	title = {Recovery of the low- and high-mass end slopes of the {IMF} in massive early-type galaxies using detailed elemental abundances},
	volume = {530},
	copyright = {https://creativecommons.org/licenses/by/4.0/},
	issn = {0035-8711, 1365-2966},
	url = {https://academic.oup.com/mnras/article/530/3/3278/7638210},
	doi = {10.1093/mnras/stae912},
	language = {en},
	number = {3},
	urldate = {2024-10-12},
	journal = {Monthly Notices of the Royal Astronomical Society},
	author = {den Brok, Mark and Krajnović, Davor and Emsellem, Eric and Mercier, Wilfried and Steinmetz, Matthias and Weilbacher, Peter M},
	month = apr,
	year = {2024},
	pages = {3278--3301},
}

@misc{gountanis_modeling_2024,
	title = {Modeling the {Ages} and {Chemical} {Abundances} of {Elliptical} {Galaxies}},
	copyright = {Creative Commons Attribution 4.0 International},
	url = {https://arxiv.org/abs/2407.07971},
	doi = {10.48550/ARXIV.2407.07971},
	urldate = {2024-09-15},
	publisher = {arXiv},
	author = {Gountanis, Nicole Marcelina and Weinberg, David H. and Beverage, Aliza G. and Sandford, Nathan R. and Conroy, Charlie and Kriek, Mariska},
	year = {2024},
	note = {Version Number: 1},
}

@article{roberts_nature_2024,
	title = {Nature versus nurture: distinguishing effects from stellar processing and chemical evolution on carbon and nitrogen in red giant stars},
	volume = {530},
	copyright = {https://creativecommons.org/licenses/by/4.0/},
	issn = {0035-8711, 1365-2966},
	shorttitle = {Nature versus nurture},
	url = {https://academic.oup.com/mnras/article/530/1/149/7636511},
	doi = {10.1093/mnras/stae820},
	language = {en},
	number = {1},
	urldate = {2024-09-15},
	journal = {Monthly Notices of the Royal Astronomical Society},
	author = {Roberts, John D and Pinsonneault, Marc H and Johnson, Jennifer A and Zinn, Joel C and Weinberg, David H and Vrard, Mathieu and Tayar, Jamie and Stello, Dennis and Mosser, Benoît and Johnson, James W and Cao, Kaili and Stassun, Keivan G and Stringfellow, Guy S and Serenelli, Aldo and Mathur, Savita and Hekker, Saskia and García, Rafael A and Elsworth, Yvonne P and Corsaro, Enrico},
	month = apr,
	year = {2024},
	pages = {149--166},
}

@article{griffith_residual_2022,
	title = {Residual {Abundances} in {GALAH} {DR3}: {Implications} for {Nucleosynthesis} and {Identification} of {Unique} {Stellar} {Populations}},
	volume = {931},
	issn = {0004-637X, 1538-4357},
	shorttitle = {Residual {Abundances} in {GALAH} {DR3}},
	url = {https://iopscience.iop.org/article/10.3847/1538-4357/ac5826},
	doi = {10.3847/1538-4357/ac5826},
	number = {1},
	urldate = {2024-09-15},
	journal = {The Astrophysical Journal},
	author = {Griffith, Emily J. and Weinberg, David H. and Buder, Sven and Johnson, Jennifer A. and Johnson, James W. and Vincenzo, Fiorenzo},
	month = may,
	year = {2022},
	pages = {23},
}

@article{jafariyazani_chemical_2024,
	title = {Chemical {Abundances} of {Early} {Quiescent} {Galaxies}: {New} {Observations} and {Modelling} {Impacts}},
	copyright = {Creative Commons Attribution 4.0 International},
	shorttitle = {Chemical {Abundances} of {Early} {Quiescent} {Galaxies}},
	url = {https://arxiv.org/abs/2406.03549},
	doi = {10.48550/ARXIV.2406.03549},
	urldate = {2024-06-12},
	author = {Jafariyazani, Marziye and Newman, Andrew B. and Mobasher, Bahram and Belli, Sirio and Ellis, Richard S. and Faisst, Andreas L.},
	year = {2024},
	note = {Publisher: [object Object]
Version Number: 1},
}

@article{carnall_jwst_2024,
	title = {The {JWST} {EXCELS} survey: {Too} much, too young, too fast? {Ultra}-massive quiescent galaxies at 3 \&lt; z \&lt; 5},
	copyright = {Creative Commons Attribution 4.0 International},
	shorttitle = {The {JWST} {EXCELS} survey},
	url = {https://arxiv.org/abs/2405.02242},
	doi = {10.48550/ARXIV.2405.02242},
	urldate = {2024-06-07},
	author = {Carnall, A. C. and Cullen, F. and McLure, R. J. and McLeod, D. J. and Begley, R. and Donnan, C. T. and Dunlop, J. S. and Shapley, A. E. and Rowlands, K. and Almaini, O. and Arellano-Córdova, K. Z. and Barrufet, L. and Cimatti, A. and Ellis, R. S. and Grogin, N. A. and Hamadouche, M. L. and Illingworth, G. D. and Koekemoer, A. M. and Leung, H. -H. and Lovell, C. C. and Pérez-González, P. G. and Santini, P. and Stanton, T. M. and Wild, V.},
	year = {2024},
	note = {Publisher: [object Object]
Version Number: 1},
}

@article{glazebrook_massive_2024,
	title = {A massive galaxy that formed its stars at z ≈ 11},
	volume = {628},
	issn = {0028-0836, 1476-4687},
	url = {https://www.nature.com/articles/s41586-024-07191-9},
	doi = {10.1038/s41586-024-07191-9},
	language = {en},
	number = {8007},
	urldate = {2024-06-07},
	journal = {Nature},
	author = {Glazebrook, Karl and Nanayakkara, Themiya and Schreiber, Corentin and Lagos, Claudia and Kawinwanichakij, Lalitwadee and Jacobs, Colin and Chittenden, Harry and Brammer, Gabriel and Kacprzak, Glenn G. and Labbe, Ivo and Marchesini, Danilo and Marsan, Z. Cemile and Oesch, Pascal A. and Papovich, Casey and Remus, Rhea-Silvia and Tran, Kim-Vy H. and Esdaile, James and Chandro-Gomez, Angel},
	month = apr,
	year = {2024},
	pages = {277--281},
}

@article{cristallo_evolution_2011,
	title = {{EVOLUTION}, {NUCLEOSYNTHESIS}, {AND} {YIELDS} {OF} {LOW}-{MASS} {ASYMPTOTIC} {GIANT} {BRANCH} {STARS} {AT} {DIFFERENT} {METALLICITIES}. {II}. {THE} {FRUITY} {DATABASE}},
	volume = {197},
	issn = {0067-0049, 1538-4365},
	url = {https://iopscience.iop.org/article/10.1088/0067-0049/197/2/17},
	doi = {10.1088/0067-0049/197/2/17},
	number = {2},
	urldate = {2024-05-16},
	journal = {The Astrophysical Journal Supplement Series},
	author = {Cristallo, S. and Piersanti, L. and Straniero, O. and Gallino, R. and Domínguez, I. and Abia, C. and Di Rico, G. and Quintini, M. and Bisterzo, S.},
	month = dec,
	year = {2011},
	pages = {17},
}

@article{cristallo_evolution_2015,
	title = {{EVOLUTION}, {NUCLEOSYNTHESIS}, {AND} {YIELDS} {OF} {AGB} {STARS} {AT} {DIFFERENT} {METALLICITIES}. {III}. {INTERMEDIATE}-{MASS} {MODELS}, {REVISED} {LOW}-{MASS} {MODELS}, {AND} {THE} {pH}-{FRUITY} {INTERFACE}},
	volume = {219},
	copyright = {http://iopscience.iop.org/info/page/text-and-data-mining},
	issn = {1538-4365},
	url = {https://iopscience.iop.org/article/10.1088/0067-0049/219/2/40},
	doi = {10.1088/0067-0049/219/2/40},
	number = {2},
	urldate = {2024-05-16},
	journal = {The Astrophysical Journal Supplement Series},
	author = {Cristallo, S. and Straniero, O. and Piersanti, L. and Gobrecht, D.},
	month = aug,
	year = {2015},
	pages = {40},
}

@article{beverage_heavy_2024,
	title = {The {Heavy} {Metal} {Survey}: {The} {Evolution} of {Stellar} {Metallicities}, {Abundance} {Ratios}, and {Ages} of {Massive} {Quiescent} {Galaxies} since z ∼ 2},
	volume = {966},
	issn = {0004-637X, 1538-4357},
	shorttitle = {The {Heavy} {Metal} {Survey}},
	url = {https://iopscience.iop.org/article/10.3847/1538-4357/ad372d},
	doi = {10.3847/1538-4357/ad372d},
	number = {2},
	urldate = {2024-05-12},
	journal = {The Astrophysical Journal},
	author = {Beverage, Aliza G. and Kriek, Mariska and Suess, Katherine A. and Conroy, Charlie and Price, Sedona H. and Barro, Guillermo and Bezanson, Rachel and Franx, Marijn and Lorenz, Brian and Ma, Yilun and Mowla, Lamiya A. and Pasha, Imad and Van Dokkum, Pieter and Weisz, Daniel R.},
	month = may,
	year = {2024},
	pages = {234},
}

@misc{slob_jwst-suspense_2024,
	title = {The {JWST}-{SUSPENSE} {Ultradeep} {Spectroscopic} {Program}: {Survey} {Overview} and {Star}-{Formation} {Histories} of {Quiescent} {Galaxies} at 1 {\textless} z {\textless} 3},
	shorttitle = {The {JWST}-{SUSPENSE} {Ultradeep} {Spectroscopic} {Program}},
	url = {http://arxiv.org/abs/2404.12432},
	language = {en},
	urldate = {2024-04-25},
	publisher = {arXiv},
	author = {Slob, Martje and Kriek, Mariska and Beverage, Aliza G. and Suess, Katherine A. and Barro, Guillermo and Bezanson, Rachel and Cheng, Chloe M. and Conroy, Charlie and de Graaff, Anna and Schreiber, Natascha M. Förster and Franx, Marijn and Lorenz, Brian and Piña, Pavel E. Mancera and Marchesini, Danilo and Muzzin, Adam and Newman, Andrew B. and Price, Sedona H. and Shapley, Alice E. and Stefanon, Mauro and van Dokkum, Pieter and Weisz, Daniel R.},
	month = apr,
	year = {2024},
	note = {arXiv:2404.12432 [astro-ph]},
}

@article{maoz_supernova_2010,
	title = {{THE} {SUPERNOVA} {DELAY} {TIME} {DISTRIBUTION} {IN} {GALAXY} {CLUSTERS} {AND} {IMPLICATIONS} {FOR} {TYPE}-{Ia} {PROGENITORS} {AND} {METAL} {ENRICHMENT}},
	volume = {722},
	issn = {0004-637X, 1538-4357},
	url = {https://iopscience.iop.org/article/10.1088/0004-637X/722/2/1879},
	doi = {10.1088/0004-637X/722/2/1879},
	language = {en},
	number = {2},
	urldate = {2024-04-15},
	journal = {The Astrophysical Journal},
	author = {Maoz, Dan and Sharon, Keren and {Avishay Gal-Yam}},
	month = oct,
	year = {2010},
	pages = {1879--1894},
}

@article{kriek_massive_2016,
	title = {A massive, quiescent, population {II} galaxy at a redshift of 2.1},
	volume = {540},
	issn = {0028-0836, 1476-4687},
	url = {http://arxiv.org/abs/1612.02001},
	doi = {10.1038/nature20570},
	number = {7632},
	urldate = {2020-11-04},
	journal = {Nature},
	author = {Kriek, Mariska and Conroy, Charlie and van Dokkum, Pieter G. and Shapley, Alice E. and Choi, Jieun and Reddy, Naveen A. and Siana, Brian and van de Voort, Freeke and Coil, Alison L. and Mobasher, Bahram},
	month = dec,
	year = {2016},
	note = {arXiv: 1612.02001},
	pages = {248--251},
}

@article{speagle_dynesty_2020,
	title = {dynesty: {A} {Dynamic} {Nested} {Sampling} {Package} for {Estimating} {Bayesian} {Posteriors} and {Evidences}},
	volume = {493},
	issn = {0035-8711, 1365-2966},
	shorttitle = {dynesty},
	url = {http://arxiv.org/abs/1904.02180},
	doi = {10.1093/mnras/staa278},
	number = {3},
	urldate = {2020-10-28},
	journal = {Monthly Notices of the Royal Astronomical Society},
	author = {Speagle, Joshua S.},
	month = apr,
	year = {2020},
	note = {arXiv: 1904.02180},
	pages = {3132--3158},
}

@article{choi_assembly_2014,
	title = {{THE} {ASSEMBLY} {HISTORIES} {OF} {QUIESCENT} {GALAXIES} {SINCE} \textit{z} = 0.7 {FROM} {ABSORPTION} {LINE} {SPECTROSCOPY}},
	volume = {792},
	issn = {1538-4357},
	url = {https://iopscience.iop.org/article/10.1088/0004-637X/792/2/95},
	doi = {10.1088/0004-637X/792/2/95},
	language = {en},
	number = {2},
	urldate = {2020-10-26},
	journal = {The Astrophysical Journal},
	author = {Choi, Jieun and Conroy, Charlie and Moustakas, John and Graves, Genevieve J. and Holden, Bradford P. and Brodwin, Mark and Brown, Michael J. I. and van Dokkum, Pieter G.},
	month = aug,
	year = {2014},
	pages = {95},
}

@article{jafariyazani_resolved_2020,
	title = {Resolved {Multi}-element {Stellar} {Chemical} {Abundances} in the {Brightest} {Quiescent} {Galaxy} at z \${\textbackslash}sim\$ 2},
	volume = {897},
	issn = {2041-8213},
	url = {http://arxiv.org/abs/2007.00205},
	doi = {10.3847/2041-8213/aba11c},
	number = {2},
	urldate = {2020-10-21},
	journal = {The Astrophysical Journal},
	author = {Jafariyazani, Marziye and Newman, Andrew B. and Mobasher, Bahram and Belli, Sirio and Ellis, Richard S. and Patel, Shannon G.},
	month = jul,
	year = {2020},
	note = {arXiv: 2007.00205},
	pages = {L42},
}

@article{mcdermid_atlas3d_2015,
	title = {The {ATLAS3D} {Project} – {XXX}. {Star} formation histories and stellar population scaling relations of early-type galaxies},
	volume = {448},
	issn = {1365-2966, 0035-8711},
	url = {http://academic.oup.com/mnras/article/448/4/3484/955288/The-ATLAS3D-Project-XXX-Star-formation-histories},
	doi = {10.1093/mnras/stv105},
	language = {en},
	number = {4},
	urldate = {2020-10-11},
	journal = {Monthly Notices of the Royal Astronomical Society},
	author = {McDermid, Richard M. and Alatalo, Katherine and Blitz, Leo and Bournaud, Frédéric and Bureau, Martin and Cappellari, Michele and Crocker, Alison F. and Davies, Roger L. and Davis, Timothy A. and de Zeeuw, P. T. and Duc, Pierre-Alain and Emsellem, Eric and Khochfar, Sadegh and Krajnović, Davor and Kuntschner, Harald and Morganti, Raffaella and Naab, Thorsten and Oosterloo, Tom and Sarzi, Marc and Scott, Nicholas and Serra, Paolo and Weijmans, Anne-Marie and Young, Lisa M.},
	month = apr,
	year = {2015},
	pages = {3484--3513},
}

@article{nomoto_nucleosynthesis_2013,
	title = {Nucleosynthesis in {Stars} and the {Chemical} {Enrichment} of {Galaxies}},
	volume = {51},
	issn = {0066-4146, 1545-4282},
	url = {http://www.annualreviews.org/doi/10.1146/annurev-astro-082812-140956},
	doi = {10.1146/annurev-astro-082812-140956},
	language = {en},
	number = {1},
	urldate = {2020-10-11},
	journal = {Annual Review of Astronomy and Astrophysics},
	author = {Nomoto, Ken'ichi and Kobayashi, Chiaki and Tominaga, Nozomu},
	month = aug,
	year = {2013},
	pages = {457--509},
}

@article{beverage_elemental_2021,
	title = {Elemental {Abundances} and {Ages} of z ∼ 0.7 {Quiescent} {Galaxies} on the {Mass}–{Size} {Plane}: {Implication} for {Chemical} {Enrichment} and {Star} {Formation} {Quenching}},
	volume = {917},
	issn = {2041-8205, 2041-8213},
	shorttitle = {Elemental {Abundances} and {Ages} of z ∼ 0.7 {Quiescent} {Galaxies} on the {Mass}–{Size} {Plane}},
	url = {https://iopscience.iop.org/article/10.3847/2041-8213/ac12cd},
	doi = {10.3847/2041-8213/ac12cd},
	language = {en},
	number = {1},
	urldate = {2021-09-27},
	journal = {The Astrophysical Journal Letters},
	author = {Beverage, Aliza G. and Kriek, Mariska and Conroy, Charlie and Bezanson, Rachel and Franx, Marijn and van der Wel, Arjen},
	month = aug,
	year = {2021},
	pages = {L1},
}

@article{conroy_strontium_2013,
	title = {{STRONTIUM} {AND} {BARIUM} {IN} {EARLY}-{TYPE} {GALAXIES}},
	volume = {763},
	issn = {2041-8205, 2041-8213},
	url = {https://iopscience.iop.org/article/10.1088/2041-8205/763/2/L25},
	doi = {10.1088/2041-8205/763/2/L25},
	language = {en},
	number = {2},
	urldate = {2021-09-14},
	journal = {The Astrophysical Journal},
	author = {Conroy, Charlie and van Dokkum, Pieter G. and Graves, Genevieve J.},
	month = jan,
	year = {2013},
	pages = {L25},
}

@article{conroy_counting_2012,
	title = {{COUNTING} {LOW}-{MASS} {STARS} {IN} {INTEGRATED} {LIGHT}},
	volume = {747},
	issn = {0004-637X, 1538-4357},
	url = {https://iopscience.iop.org/article/10.1088/0004-637X/747/1/69},
	doi = {10.1088/0004-637X/747/1/69},
	language = {en},
	number = {1},
	urldate = {2021-08-19},
	journal = {The Astrophysical Journal},
	author = {Conroy, Charlie and van Dokkum, Pieter},
	month = mar,
	year = {2012},
	pages = {69},
}

@article{weinberg_chemical_2019,
	title = {Chemical {Cartography} with {APOGEE}: {Multi}-element abundance ratios},
	volume = {874},
	issn = {1538-4357},
	shorttitle = {Chemical {Cartography} with {APOGEE}},
	url = {http://arxiv.org/abs/1810.12325},
	doi = {10.3847/1538-4357/ab07c7},
	number = {1},
	urldate = {2021-06-25},
	journal = {The Astrophysical Journal},
	author = {Weinberg, David H. and Holtzman, Jon A. and Hasselquist, Sten and Bird, Jonathan C. and Johnson, Jennifer A. and Shetrone, Matthew and Sobeck, Jennifer and Prieto, Carlos Allende and Bizyaev, Dmitry and Carrera, Ricardo and Cohen, Roger E. and Cunha, Katia and Ebelke, Garrett and Fernandez-Trincado, J. G. and Garcia-Hernandez, D. A. and Hayes, Christian R. and Jonsson, Henrik and Lane, Richard R. and Majewski, Steven R. and Malanushenko, Viktor and Meszaros, Szabolcz and Nidever, David L. and Nitschelm, Christian and Pan, Kaike and Schiavon, Ricardo P. and Schneider, Donald P. and Wilson, John C. and Zamora, Olga},
	month = mar,
	year = {2019},
	note = {arXiv: 1810.12325},
	pages = {102},
}

@article{foreman-mackey_emcee_2013,
	title = {emcee : {The} {MCMC} {Hammer}},
	volume = {125},
	issn = {00046280, 15383873},
	shorttitle = {emcee},
	url = {http://iopscience.iop.org/article/10.1086/670067},
	doi = {10.1086/670067},
	language = {en},
	number = {925},
	urldate = {2020-12-11},
	journal = {Publications of the Astronomical Society of the Pacific},
	author = {Foreman-Mackey, Daniel and Hogg, David W. and Lang, Dustin and Goodman, Jonathan},
	month = mar,
	year = {2013},
	pages = {306--312},
}

@article{villaume_extended_2017,
	title = {The {Extended} {IRTF} {Spectral} {Library}: {Expanded} {Coverage} in {Metallicity}, {Temperature}, and {Surface} {Gravity}},
	volume = {230},
	issn = {1538-4365},
	shorttitle = {The {Extended} {IRTF} {Spectral} {Library}},
	url = {https://iopscience.iop.org/article/10.3847/1538-4365/aa72ed},
	doi = {10.3847/1538-4365/aa72ed},
	language = {en},
	number = {2},
	urldate = {2020-12-11},
	journal = {The Astrophysical Journal Supplement Series},
	author = {Villaume, Alexa and Conroy, Charlie and Johnson, Benjamin and Rayner, John and Mann, Andrew W. and van Dokkum, Pieter},
	month = jun,
	year = {2017},
	pages = {23},
}

@article{sanchez-blazquez_medium-resolution_2006,
	title = {Medium-resolution {Isaac} {Newton} {Telescope} library of empirical spectra},
	volume = {371},
	issn = {0035-8711, 1365-2966},
	url = {https://academic.oup.com/mnras/article-lookup/doi/10.1111/j.1365-2966.2006.10699.x},
	doi = {10.1111/j.1365-2966.2006.10699.x},
	language = {en},
	number = {2},
	urldate = {2020-12-11},
	journal = {Monthly Notices of the Royal Astronomical Society},
	author = {Sanchez-Blazquez, P. and Peletier, R. F. and Jimenez-Vicente, J. and Cardiel, N. and Cenarro, A. J. and Falcon-Barroso, J. and Gorgas, J. and Selam, S. and Vazdekis, A.},
	month = sep,
	year = {2006},
	pages = {703--718},
}

@article{choi_mesa_2016,
	title = {{MESA} {ISOCHRONES} {AND} {STELLAR} {TRACKS} ({MIST}). {I}. {SOLAR}-{SCALED} {MODELS}},
	volume = {823},
	issn = {1538-4357},
	url = {https://iopscience.iop.org/article/10.3847/0004-637X/823/2/102},
	doi = {10.3847/0004-637X/823/2/102},
	language = {en},
	number = {2},
	urldate = {2020-12-11},
	journal = {The Astrophysical Journal},
	author = {Choi, Jieun and Dotter, Aaron and Conroy, Charlie and Cantiello, Matteo and Paxton, Bill and Johnson, Benjamin D.},
	month = may,
	year = {2016},
	pages = {102},
}

@article{conroy_early-type_2014,
	title = {{EARLY}-{TYPE} {GALAXY} {ARCHEOLOGY}: {AGES}, {ABUNDANCE} {RATIOS}, {AND} {EFFECTIVE} {TEMPERATURES} {FROM} {FULL}-{SPECTRUM} {FITTING}},
	volume = {780},
	issn = {0004-637X, 1538-4357},
	shorttitle = {{EARLY}-{TYPE} {GALAXY} {ARCHEOLOGY}},
	url = {https://iopscience.iop.org/article/10.1088/0004-637X/780/1/33},
	doi = {10.1088/0004-637X/780/1/33},
	language = {en},
	number = {1},
	urldate = {2020-12-11},
	journal = {The Astrophysical Journal},
	author = {Conroy, Charlie and Graves, Genevieve J. and van Dokkum, Pieter G.},
	month = jan,
	year = {2014},
	pages = {33},
}

@article{van_der_wel_vlt_2016,
	title = {{THE} {VLT} {LEGA}-{C} {SPECTROSCOPIC} {SURVEY}: {THE} {PHYSICS} {OF} {GALAXIES} {AT} {A} {LOOKBACK} {TIME} {OF} 7 {Gyr}},
	volume = {223},
	issn = {1538-4365},
	shorttitle = {{THE} {VLT} {LEGA}-{C} {SPECTROSCOPIC} {SURVEY}},
	url = {https://iopscience.iop.org/article/10.3847/0067-0049/223/2/29},
	doi = {10.3847/0067-0049/223/2/29},
	language = {en},
	number = {2},
	urldate = {2020-12-11},
	journal = {The Astrophysical Journal Supplement Series},
	author = {van der Wel, A. and Noeske, K. and Bezanson, R. and Pacifici, C. and Gallazzi, A. and Franx, M. and Muñoz-Mateos, J. C. and Bell, E. F. and Brammer, G. and Charlot, S. and Chauké, P. and Labbé, I. and Maseda, M. V. and Muzzin, A. and Rix, H.-W. and Sobral, D. and Sande, J. van de and Dokkum, P. G. van and Wild, V. and Wolf, C.},
	month = apr,
	year = {2016},
	pages = {29},
}

@article{straatman_large_2018,
	title = {The {Large} {Early} {Galaxy} {Astrophysics} {Census} ({LEGA}-{C}) {Data} {Release} 2: {Dynamical} and {Stellar} {Population} {Properties} of \textit{z} ≲ 1 {Galaxies} in the {COSMOS} {Field}},
	volume = {239},
	issn = {1538-4365},
	shorttitle = {The {Large} {Early} {Galaxy} {Astrophysics} {Census} ({LEGA}-{C}) {Data} {Release} 2},
	url = {https://iopscience.iop.org/article/10.3847/1538-4365/aae37a},
	doi = {10.3847/1538-4365/aae37a},
	language = {en},
	number = {2},
	urldate = {2020-12-11},
	journal = {The Astrophysical Journal Supplement Series},
	author = {Straatman, Caroline M. S. and Wel, Arjen van der and Bezanson, Rachel and Pacifici, Camilla and Gallazzi, Anna and Wu, Po-Feng and Noeske, Kai and Barišić, Ivana and Bell, Eric F. and Brammer, Gabriel B. and Calhau, João and Chauke, Priscilla and Franx, Marijn and Houdt, Josha van and Labbé, Ivo and Maseda, Michael V. and Muñoz-Mateos, Juan C. and Muzzin, Adam and Sande, Jesse van de and Sobral, David and Spilker, Justin S.},
	month = dec,
	year = {2018},
	pages = {27},
}

@article{muzzin_public_2013,
	title = {A {PUBLIC} \textit{ {K} $_{\textrm{s}}$ } -{SELECTED} {CATALOG} {IN} {THE} {COSMOS}/{ULTRAVISTA} {FIELD}: {PHOTOMETRY}, {PHOTOMETRIC} {REDSHIFTS}, {AND} {STELLAR} {POPULATION} {PARAMETERS} $^{\textrm{,}}$},
	volume = {206},
	issn = {0067-0049, 1538-4365},
	shorttitle = {A {PUBLIC} \textit{ {K} $_{\textrm{s}}$ } -{SELECTED} {CATALOG} {IN} {THE} {COSMOS}/{ULTRAVISTA} {FIELD}},
	url = {https://iopscience.iop.org/article/10.1088/0067-0049/206/1/8},
	doi = {10.1088/0067-0049/206/1/8},
	language = {en},
	number = {1},
	urldate = {2020-12-11},
	journal = {The Astrophysical Journal Supplement Series},
	author = {Muzzin, Adam and Marchesini, Danilo and Stefanon, Mauro and Franx, Marijn and Milvang-Jensen, Bo and Dunlop, James S. and Fynbo, J. P. U. and Brammer, Gabriel and Labbé, Ivo and van Dokkum, Pieter},
	month = may,
	year = {2013},
	pages = {8},
}

@article{conroy_metal-rich_2018,
	title = {Metal-rich, {Metal}-poor: {Updated} {Stellar} {Population} {Models} for {Old} {Stellar} {Systems}},
	volume = {854},
	issn = {1538-4357},
	shorttitle = {Metal-rich, {Metal}-poor},
	url = {http://arxiv.org/abs/1801.10185},
	doi = {10.3847/1538-4357/aaab49},
	number = {2},
	urldate = {2020-12-11},
	journal = {The Astrophysical Journal},
	author = {Conroy, Charlie and Villaume, Alexa and van Dokkum, Pieter and Lind, Karin},
	month = feb,
	year = {2018},
	note = {arXiv: 1801.10185},
	pages = {139},
}

@article{van_der_wel_large_2021,
	title = {The {Large} {Early} {Galaxy} {Astrophysics} {Census} ({LEGA}-{C}) {Data} {Release} 3: 3000 {High}-quality {Spectra} of {K} $_{\textrm{s}}$ -selected {Galaxies} at z {\textgreater} 0.6},
	volume = {256},
	issn = {0067-0049, 1538-4365},
	shorttitle = {The {Large} {Early} {Galaxy} {Astrophysics} {Census} ({LEGA}-{C}) {Data} {Release} 3},
	url = {https://iopscience.iop.org/article/10.3847/1538-4365/ac1356},
	doi = {10.3847/1538-4365/ac1356},
	language = {en},
	number = {2},
	urldate = {2022-05-27},
	journal = {The Astrophysical Journal Supplement Series},
	author = {van der Wel, Arjen and Bezanson, Rachel and D’Eugenio, Francesco and Straatman, Caroline and Franx, Marijn and van Houdt, Josha and Maseda, Michael V. and Gallazzi, Anna and Wu, Po-Feng and Pacifici, Camilla and Barisic, Ivana and Brammer, Gabriel B. and Munoz-Mateos, Juan Carlos and Vervalcke, Sarah and Zibetti, Stefano and Sobral, David and de Graaff, Anna and Calhau, Joao and Kaushal, Yasha and Muzzin, Adam and Bell, Eric F. and van Dokkum, Pieter G.},
	month = oct,
	year = {2021},
	pages = {44},
}

@article{graves_dissecting_2010,
	title = {{DISSECTING} {THE} {RED} {SEQUENCE}. {III}. {MASS}-{TO}-{LIGHT} {VARIATIONS} {IN} {THREE}-{DIMENSIONAL} {FUNDAMENTAL} {PLANE} {SPACE}},
	volume = {717},
	issn = {0004-637X, 1538-4357},
	url = {https://iopscience.iop.org/article/10.1088/0004-637X/717/2/803},
	doi = {10.1088/0004-637X/717/2/803},
	language = {en},
	number = {2},
	urldate = {2022-08-30},
	journal = {The Astrophysical Journal},
	author = {Graves, Genevieve J. and Faber, S. M.},
	month = jul,
	year = {2010},
	pages = {803--824},
}

@article{trager_stellar_2000,
	title = {The {Stellar} {Population} {Histories} of {Early}-{Type} {Galaxies}. {II}. {Controlling} {Parameters} of the {Stellar} {Populations}},
	volume = {120},
	issn = {00046256},
	url = {https://iopscience.iop.org/article/10.1086/301442},
	doi = {10.1086/301442},
	language = {en},
	number = {1},
	urldate = {2022-08-30},
	journal = {The Astronomical Journal},
	author = {Trager, S. C. and Faber, S. M. and Worthey, Guy and González, J. Jesús},
	month = jul,
	year = {2000},
	pages = {165--188},
}

@article{thomas_environment_2010-1,
	title = {Environment and self-regulation in galaxy formation},
	issn = {00358711, 13652966},
	url = {https://academic.oup.com/mnras/article-lookup/doi/10.1111/j.1365-2966.2010.16427.x},
	doi = {10.1111/j.1365-2966.2010.16427.x},
	language = {en},
	urldate = {2022-08-30},
	journal = {Monthly Notices of the Royal Astronomical Society},
	author = {Thomas, Daniel and Maraston, Claudia and Schawinski, Kevin and Sarzi, Marc and Silk, Joseph},
	month = mar,
	year = {2010},
}

@article{matteucci_abundance_1994,
	title = {Abundance ratios in ellipticals and galaxy formation.},
	volume = {288},
	issn = {0004-6361},
	url = {https://ui.adsabs.harvard.edu/abs/1994A&A...288...57M},
	urldate = {2022-08-30},
	journal = {Astronomy and Astrophysics},
	author = {Matteucci, Francesca},
	month = aug,
	year = {1994},
	note = {ADS Bibcode: 1994A\&A...288...57M},
	pages = {57--64},
}

@article{thomas_epochs_2005,
	title = {{THE} {EPOCHS} {OF} {EARLY}-{TYPE} {GALAXY} {FORMATION} {AS} {A} {FUNCTION} {OF} {ENVIRONMENT}},
	volume = {621},
	doi = {10.1086/426932},
	language = {en},
	number = {2},
	journal = {The Astronomical Journal},
	author = {Thomas, Daniel and Maraston, Claudia and Bender, Ralf},
	month = mar,
	year = {2005},
	pages = {22},
}

@article{tinsley_stellar_1979,
	title = {Stellar lifetimes and abundance ratios in chemical evolution},
	volume = {229},
	issn = {0004-637X, 1538-4357},
	url = {http://adsabs.harvard.edu/doi/10.1086/157039},
	doi = {10.1086/157039},
	language = {en},
	urldate = {2022-10-07},
	journal = {The Astrophysical Journal},
	author = {Tinsley, B. M.},
	month = may,
	year = {1979},
	pages = {1046},
}

@article{nomoto_nucleosynthesis_2006,
	title = {Nucleosynthesis {Yields} of {Core}-{Collapse} {Supernovae} and {Hypernovae}, and {Galactic} {Chemical} {Evolution}},
	volume = {777},
	issn = {03759474},
	url = {http://arxiv.org/abs/astro-ph/0605725},
	doi = {10.1016/j.nuclphysa.2006.05.008},
	urldate = {2022-11-03},
	journal = {Nuclear Physics A},
	author = {Nomoto, Ken'ichi and Tominaga, Nozomu and Umeda, Hideyuki and Kobayashi, Chiaki and Maeda, Keiichi},
	month = oct,
	year = {2006},
	note = {arXiv:astro-ph/0605725},
	pages = {424--458},
}

@article{johnson_empirical_2023,
	title = {Empirical constraints on the nucleosynthesis of nitrogen},
	volume = {520},
	issn = {0035-8711, 1365-2966},
	url = {https://academic.oup.com/mnras/article/520/1/782/7005237},
	doi = {10.1093/mnras/stad057},
	language = {en},
	number = {1},
	urldate = {2023-03-22},
	journal = {Monthly Notices of the Royal Astronomical Society},
	author = {Johnson, James W and Weinberg, David H and Vincenzo, Fiorenzo and Bird, Jonathan C and Griffith, Emily J},
	month = jan,
	year = {2023},
	pages = {782--803},
}

@article{beverage_carbon_2023,
	title = {From {Carbon} to {Cobalt}: {Chemical} {Compositions} and {Ages} of z ∼ 0.7 {Quiescent} {Galaxies}},
	volume = {948},
	issn = {0004-637X},
	shorttitle = {From {Carbon} to {Cobalt}},
	url = {https://dx.doi.org/10.3847/1538-4357/acc176},
	doi = {10.3847/1538-4357/acc176},
	language = {en},
	number = {2},
	urldate = {2023-12-08},
	journal = {The Astrophysical Journal},
	author = {Beverage, Aliza G. and Kriek, Mariska and Conroy, Charlie and Sandford, Nathan R. and Bezanson, Rachel and Franx, Marijn and Wel, Arjen van der and Weisz, Daniel R.},
	month = may,
	year = {2023},
	note = {Publisher: The American Astronomical Society},
	pages = {140},
}

@article{zhuang_glimpse_2023,
	title = {A {Glimpse} of the {Stellar} {Populations} and {Elemental} {Abundances} of {Gravitationally} {Lensed}, {Quiescent} {Galaxies} at z ≳ 1 with {Keck} {Deep} {Spectroscopy}},
	volume = {948},
	issn = {0004-637X, 1538-4357},
	url = {https://iopscience.iop.org/article/10.3847/1538-4357/acc79b},
	doi = {10.3847/1538-4357/acc79b},
	language = {en},
	number = {2},
	urldate = {2023-06-07},
	journal = {The Astrophysical Journal},
	author = {Zhuang, Zhuyun and Leethochawalit, Nicha and Kirby, Evan N. and Nightingale, J. W. and Steidel, Charles C. and Glazebrook, Karl and Barone, Tania M. and Skobe, Hannah and Sweet, Sarah M. and Nanayakkara, Themiya and Allen, Rebecca J. and G. C., Keerthi Vasan and Jones, Tucker and Kacprzak, Glenn G. and Tran, Kim-Vy H. and Jacobs, Colin},
	month = may,
	year = {2023},
	pages = {132},
}

@article{knowles_smiles_2023,
	title = {{sMILES} {SSPs}: {A} {Library} of {Semi}-{Empirical} {MILES} {Stellar} {Population} {Models} with {Variable} [\${\textbackslash}alpha\$/{Fe}] {Abundances}},
	volume = {523},
	issn = {0035-8711, 1365-2966},
	shorttitle = {{sMILES} {SSPs}},
	url = {http://arxiv.org/abs/2306.05942},
	doi = {10.1093/mnras/stad1647},
	number = {3},
	urldate = {2023-06-28},
	journal = {Monthly Notices of the Royal Astronomical Society},
	author = {Knowles, Adam T. and Sansom, Anne E. and Vazdekis, Alex and Prieto, Carlos Allende},
	month = jun,
	year = {2023},
	note = {arXiv:2306.05942 [astro-ph]},
	pages = {3450--3470},
}

@ARTICLE{vanDokkum2024,
       author = {{van Dokkum}, Pieter and {Conroy}, Charlie},
        title = "{Reconciling M/L Ratios Across Cosmic Time: a Concordance IMF for Massive Galaxies}",
      journal = {\apjl},
         year = 2024,
        month = sep,
       volume = {973},
       number = {1},
          eid = {L32},
        pages = {L32},
          doi = {10.3847/2041-8213/ad77b8},
archivePrefix = {arXiv},
       eprint = {2407.06281},
 primaryClass = {astro-ph.GA},
       adsurl = {https://ui.adsabs.harvard.edu/abs/2024ApJ...973L..32V}
}

@article{Sukhbold2016,
    author = {Sukhbold, Tuguldur and Ertl, Thomas and Woosley, S. E. and Brown, J. M. and Janka, H.-T.},
    title = {Core-collapse Supernovae from 9 to 120 Solar Masses Based on Neutrino-powered Explosions},
    journal = {Astrophysical Journal},
    volume = {821},
    number = {1},
    pages = {38},
    year = {2016},
    doi = {10.3847/0004-637X/821/1/38}
}

@ARTICLE{gountanis2025,
       author = {{Gountanis}, Nicole Marcelina  and {Weinberg}, David H. and {Beverage}, Aliza G. and {Sandford}, Nathan R. and {Conroy}, Charlie and {Kriek}, Mariska},
        title = "{Modeling the Ages and Chemical Abundances of Elliptical Galaxies}",
      journal = {arXiv e-prints},
         year = 2024,
        month = jul,
          eid = {arXiv:2407.07971},
        pages = {arXiv:2407.07971},
          doi = {10.48550/arXiv.2407.07971},
archivePrefix = {arXiv},
       eprint = {2407.07971},
 primaryClass = {astro-ph.GA},
       adsurl = {https://ui.adsabs.harvard.edu/abs/2024arXiv240707971M}
}
\bibliographystyle{aasjournal}

\appendix

\section{Empirical yields}

\begin{figure*}
    \centering
    \includegraphics[width=1\textwidth]{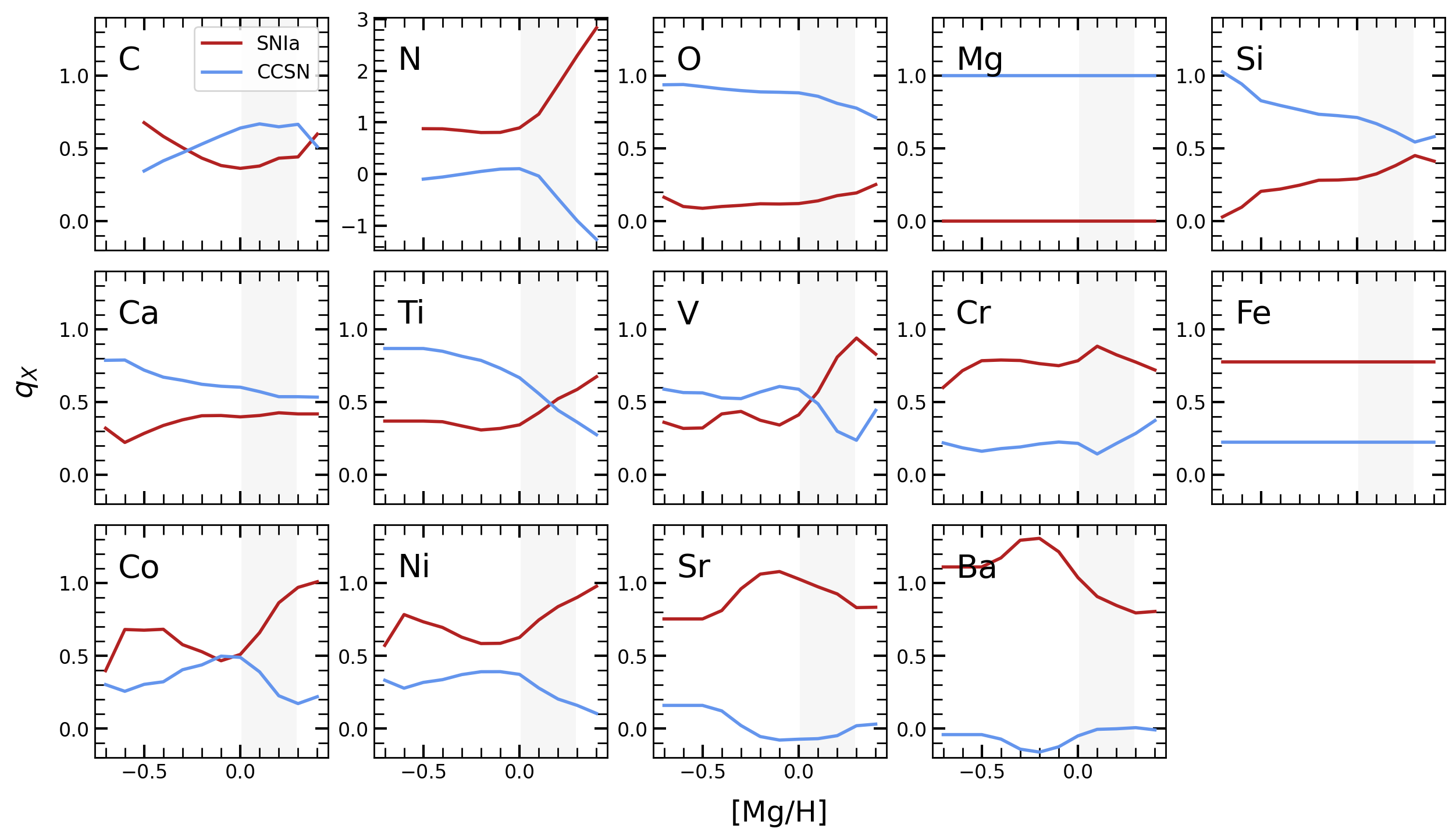}
    \caption{The adopted two-process vectors, $q^{X}_{\rm Ia}$ and $q^{X}_{\rm cc}$, adopted in this study. They are derived from the APOGEE and GALAH surveys. The quantity $q_{\rm x}$ is defined such that, at solar metallicity, the sum of $\qxcc$ and $\qxia$ for each element equals unity. All panels share the same y-axis range except for nitrogen, for which the range is expanded to highlight its strong metallicity dependence. }
    \label{fig:qs}
\end{figure*}

We present our adopted two-process vectors, $\qxcc$ and $\qxia$, together with the corresponding empirical IMF-averaged yields, $\yxcc$ and $\yxia$. Figure~\ref{fig:qs} shows $\qxcc$ and $\qxia$ as a function of metallicity. By construction, the vectors are normalized such that $\qxcc + \qxia = 1$ at solar metallicity for each element. Their relative values, therefore, quantify the fractional contributions of CCSNe and SNe Ia to the abundance of each element. These two-process vectors were derived assuming a CCSNe plateau [Mg/Fe]$_{\rm cc}=0.65$. Previous derivations of $q^X_{\rm cc}$ and $q^X_{\rm Ia}$ typically assume lower values of [Mg/Fe]$_{\rm cc}$.$^a$

We convert these two-process vectors into empirically calibrated IMF-averaged yields, $\yxcc$ and $\yxia$ in Section~\ref{sec:subtle}, and present them in Tables~\ref{tab:yields} and \ref{tab:yields_cnba}. We present C, N, Sr, and Ba separately in Table~\ref{tab:yields_cnba} because these elements should not be interpreted as physical SNe yields. These elements have a significant contribution from AGB stars, a channel not captured by the two-process decomposition (see Section~\ref{sec:agb}). These values are nonetheless used in our predictions; thus, we present them for completeness. 

The quantities reported in these tables are dimensionless and represent the mass of element X produced per unit mass of stars formed. In calculating $\yxcc$ and $\yxia$, we require the solar mass fraction of each element, $Z_{\rm X}$, which we compute following \citet{weinberg24_yields}: we adopt solar photospheric abundances, $x$ from \citet{magg22}, and convert to proto-solar abundances by adding a gravitational settling term of 0.04 dex to each element. The mass fraction can be computed using the following equation from \citetalias{weinberg22_apogee} (their Eq.~1),

\begin{equation}
    \log Z_{\rm X} = (x-12) + \log0.71 + \log A
\end{equation}

\noindent where 0.71 is the assumed solar hydrogen mass fraction and $A$ is the mean atomic weight of each element. \citet{magg22} does not include the solar photospheric abundance for Sr and Ba, and we instead adopt the values of \citet{lodders25}. 

\vspace{1em}
\noindent\parbox{\textwidth}{\footnotesize $^a$Because our assumed $\mathrm{[Mg/Fe]_{cc}}$ implies that ``high-$\alpha$ disk stars already have substantial SNe Ia enrichment, the implied values of $\qxcc$ are negative for some metallicity ranges of some elements---N, Sr, Ba---where the observed [X/Mg] gap between low- and high-$\alpha$ stars is especially large. Notably, these are also elements where the delayed contribution is expected to come from AGB stars rather than SNe Ia. While the adopted value of [Mg/Fe]$_\mathrm{cc}$ has a strong impact on the inferred $\qxcc$ and $\qxia$, we reiterate that it does \textit{not} change the predicted abundances at a given [Mg/H] and [Mg/Fe] (see Section~\ref{sec:yields})}

\begin{deluxetable*}{c|cccccccccc}
\tablecaption{Empirically calibrated IMF-averaged yields, $\yxcc$ and $\yxia$.\label{tab:yields}}
\tablehead{
\colhead{[Mg/H]} & \colhead{O} & \colhead{Mg} & \colhead{Si} & \colhead{Ca} & \colhead{Ti} & \colhead{V} & \colhead{Cr} & \colhead{Fe} & \colhead{Co} & \colhead{Ni}
}
\startdata
& & & & & & & & & & \\
CCSN, $\yxcc$ & $10^{-3}$ & $10^{-3}$ & $10^{-4}$ & $10^{-5}$ & $10^{-6}$ & $10^{-7}$ & $10^{-6}$ & $10^{-4}$ & $10^{-6}$ & $10^{-5}$ \\
\hline
-0.70 & 10.6 & 1.04 & 13.4 & 8.87 & \nodata & 2.79 & 7.53 & 4.75 & 1.91 & 4.07 \\
-0.60 & 10.6 & 1.04 & 12.3 & 8.89 & \nodata & 2.68 & 6.36 & 4.75 & 1.62 & 3.41 \\
-0.50 & 10.4 & 1.04 & 10.8 & 8.11 & 4.34 & 2.67 & 5.56 & 4.75 & 1.92 & 3.89 \\
-0.40 & 10.3 & 1.04 & 10.4 & 7.57 & 4.25 & 2.51 & 6.19 & 4.75 & 2.03 & 4.12 \\
-0.30 & 10.1 & 1.04 & 10.0 & 7.32 & 4.08 & 2.49 & 6.57 & 4.75 & 2.55 & 4.55 \\
-0.20 & 10.0 & 1.04 & 9.62 & 7.02 & 3.93 & 2.7 & 7.28 & 4.75 & 2.76 & 4.79 \\
-0.10 & 10.0 & 1.04 & 9.49 & 6.86 & 3.66 & 2.88 & 7.73 & 4.75 & 3.14 & 4.8 \\
0.00 & 9.96 & 1.04 & 9.32 & 6.79 & 3.34 & 2.79 & 7.41 & 4.75 & 3.09 & 4.57 \\
0.10 & 9.69 & 1.04 & 8.77 & 6.45 & 2.79 & 2.32 & 4.92 & 4.75 & 2.46 & 3.42 \\
0.20 & 9.14 & 1.04 & 8.01 & 6.05 & 2.22 & 1.42 & 7.4 & 4.75 & 1.43 & 2.49 \\
0.30 & 8.77 & 1.04 & 7.12 & 6.05 & 1.81 & 1.13 & 9.76 & 4.75 & 1.09 & 1.96 \\
0.40 & 8.04 & 1.04 & 7.6 & 6.02 & 1.38 & 2.1 & 12.8 & 4.75 & 1.38 & 1.28 \\
\hline
& & & & & & & & & & \\
SNIa, $\yxia$ & $10^{-4}$ & $10^{0}$ & $10^{-5}$ & $10^{-5}$ & $10^{-6}$ & $10^{-7}$ & $10^{-5}$ & $10^{-3}$ & $10^{-6}$ & $10^{-5}$ \\
\hline
-0.70 & 16.8 & 0.0 & 3.34 & 3.28 & \nodata & 1.56 & 1.87 & 1.49 & 2.28 & 6.37 \\
-0.60 & 10.2 & 0.0 & 11.3 & 2.28 & \nodata & 1.38 & 2.23 & 1.49 & 3.90 & 8.72 \\
-0.50 & 8.95 & 0.0 & 24.3 & 2.92 & 1.68 & 1.39 & 2.44 & 1.49 & 3.88 & 8.17 \\
-0.40 & 10.2 & 0.0 & 26.1 & 3.47 & 1.66 & 1.80 & 2.46 & 1.49 & 3.91 & 7.73 \\
-0.30 & 11.0 & 0.0 & 29.3 & 3.87 & 1.53 & 1.88 & 2.45 & 1.49 & 3.31 & 6.99 \\
-0.20 & 12.1 & 0.0 & 33.4 & 4.16 & 1.40 & 1.62 & 2.38 & 1.49 & 3.03 & 6.51 \\
-0.10 & 12.0 & 0.0 & 33.5 & 4.18 & 1.45 & 1.48 & 2.34 & 1.49 & 2.67 & 6.52 \\
0.00 & 12.3 & 0.0 & 34.5 & 4.08 & 1.56 & 1.78 & 2.44 & 1.49 & 2.92 & 6.97 \\
0.10 & 14.2 & 0.0 & 38.6 & 4.18 & 1.94 & 2.46 & 2.75 & 1.49 & 3.78 & 8.31 \\
0.20 & 17.9 & 0.0 & 45.4 & 4.37 & 2.38 & 3.49 & 2.57 & 1.49 & 4.95 & 9.33 \\
0.30 & 19.8 & 0.0 & 53.6 & 4.29 & 2.67 & 4.05 & 2.41 & 1.49 & 5.56 & 10.0 \\
0.40 & 25.7 & 0.0 & 48.9 & 4.29 & 3.06 & 3.58 & 2.24 & 1.49 & 5.78 & 10.9 \\
\enddata
\tablecomments{All values correspond to $[\mathrm{Mg/Fe}]_{\rm cc} = 0.65$, 
but can be rescaled to any $[\mathrm{Mg/Fe}]_{\rm cc}$. Ti yields are 
derived from GALAH abundances \citep{griffith19_galah}; all other elements 
use Milky Way disk abundances from APOGEE \citep{weinberg22_apogee}.}
\end{deluxetable*}

\begin{deluxetable*}{c|cccc|cccc}
\tablecaption{Empirical yields for elements with significant AGB enrichment.\label{tab:yields_cnba}}
\tablehead{
\multicolumn{1}{c}{} & \multicolumn{4}{c}{CCSN, $\yxcc$} & \multicolumn{4}{c}{SNIa, $\yxia$}\\ \hline
\multicolumn{1}{c|}{[Mg/H]} & \multicolumn{1}{c}{C} & \multicolumn{1}{c}{N} & \multicolumn{1}{c}{Sr} & \multicolumn{1}{c|}{Ba} & \multicolumn{1}{c}{C} & \multicolumn{1}{c}{N} & \multicolumn{1}{c}{Sr} & \multicolumn{1}{c}{Ba} \\
\multicolumn{1}{c|}{} & \multicolumn{1}{c}{$10^{-3}$} & \multicolumn{1}{c}{$10^{-5}$} & \multicolumn{1}{c}{$10^{-9}$} & \multicolumn{1}{c|}{$10^{-10}$} & \multicolumn{1}{c}{$10^{-3}$} & \multicolumn{1}{c}{$10^{-3}$} & \multicolumn{1}{c}{$10^{-8}$} & \multicolumn{1}{c}{$10^{-8}$}
}
\startdata
-0.50 & 1.80 & -15.6 & 14.3 & -12.5 & 3.21 & 1.28 & 6.13 & 3.10 \\
-0.40 & 2.16 & -8.97 & 10.9 & -21.9 & 2.76 & 1.28 & 6.60 & 3.27 \\
-0.30 & 2.46 & -0.382 & 1.95 & -43.2 & 2.40 & 1.23 & 7.82 & 3.61 \\
-0.20 & 2.77 & 8.43 & -4.84 & -49.3 & 2.06 & 1.17 & 8.64 & 3.65 \\
-0.10 & 3.06 & 15.5 & -7.00 & -38.2 & 1.81 & 1.18 & 8.78 & 3.39 \\
0.00 & 3.34 & 16.9 & -6.48 & -15.1 & 1.72 & 1.31 & 8.36 & 2.89 \\
0.10 & 3.49 & -6.09 & -6.14 & -1.39 & 1.80 & 1.69 & 7.93 & 2.53 \\
0.20 & 3.39 & -76.2 & -4.29 & -0.187 & 2.05 & 2.51 & 7.53 & 2.36 \\
0.30 & 3.48 & -145 & 1.81 & 2.17 & 2.09 & 3.35 & 6.76 & 2.22 \\
0.40 & 2.67 & -203 & 2.81 & -2.54 & 2.84 & 4.13 & 6.78 & 2.25 \\
\enddata
\tablecomments{All values correspond to $[\mathrm{Mg/Fe}]_{\rm cc} = 0.65$. C and N yields are derived from APOGEE subgiant star abundances \citep{roberts_nature_2024}; Ba yields from GALAH \citep{griffith19_galah}; and Sr yields from GALAH Y abundances as a proxy. Note that C, N, Sr, and Ba have significant AGB contributions not captured by the two-process decomposition, so these yields do not represent physical nucleosynthetic yields in the usual sense -- they are included because they are used in the model predictions of Sections~\ref{sec:mod_app_to_obs} and \ref{sec:metev}.}
\end{deluxetable*}

\end{document}